\documentclass[11pt,a4paper]{article}

\usepackage[T1]{fontenc}
\usepackage[utf8]{inputenc}
\usepackage{lmodern}
\usepackage[margin=1in]{geometry}
\usepackage{microtype}
\usepackage{graphicx}
\usepackage{amsmath,amssymb}
\usepackage{textcomp}
\usepackage{booktabs}
\usepackage{authblk}
\usepackage[font=small,labelfont=bf,labelsep=period]{caption}
\usepackage[numbers,sort&compress]{natbib}
\usepackage{xcolor}
\usepackage[
    colorlinks=true,
    linkcolor=blue,
    citecolor=blue,
    urlcolor=blue
]{hyperref}
\usepackage{setspace}

\hypersetup{
    pdftitle={Control-Noise Trade-offs in Industrially Integrated 28Si/SiGe Spin-1/2 Qubits},
    pdfauthor={Leon C. Camenzind et al.}
}

\title{\bfseries\boldmath Control--Noise Trade-offs in Industrially Integrated ${}^{28}\mathrm{Si}/\mathrm{SiGe}$ Spin-1/2 Qubits \par\vspace{1em}}

\author[1]{Leon~C.~Camenzind}
\author[1]{Ik~Kyeong~Jin}
\author[1]{Yi-Hsien~Wu}
\author[1]{Akito~Noiri}
\author[1]{Kenta~Takeda}
\author[1]{Takashi~Nakajima}
\author[2]{Takashi~Kobayashi}
\author[3]{Giordano~Scappucci}
\author[1,2]{Seigo~Tarucha}

\affil[1]{RIKEN Center for Emergent Matter Science (CEMS), RIKEN, Hirosawa 2-1, Wako-shi, 351-0198, Saitama, Japan}

\affil[2]{RIKEN Center for Quantum Computing (RQC), RIKEN, Hirosawa 2-1, Wako-shi, 351-0198, Saitama, Japan}

\affil[3]{QuTech and Kavli Institute of Nanoscience, Delft University of Technology, Lorentzweg 1, Delft, 2628~CJ, Netherlands}

\date{}

\begin{document}
\maketitle

\begingroup
\renewcommand{\thefootnote}{}
\footnotetext{\textbf{Correspondence:} L.C.C. (leon.camenzind@riken.jp) or S.T. (tarucha@riken.jp).}
\endgroup

\vspace{-2em}

\begin{abstract}
Scaling silicon spin qubits to fault-tolerant processors requires semiconductor foundry integration, yet it remains unclear how industrial fabrication and dense device integration affect qubit control and noise. Here we study ${}^{28}\mathrm{Si}/\mathrm{SiGe}$ electron spin-1/2 qubits fabricated using Intel's industry-compatible 300\,mm process and incorporating micromagnets for electrically driven spin resonance, using a high-performance device fabricated in the RIKEN research cleanroom as a reference. Primitive $\pi/2$-gate fidelities reach the 99.99\% level in the Intel devices, which also exhibit longer idle coherence, weaker low-frequency correlations and reduced qubit-to-qubit variability across several measured characteristics. In the present Intel devices, the micromagnets are integrated inline with the back-end-of-line (BEOL) contact routing, with their separation from the quantum well set through the vertical stack design. Relative to the reference, their larger separation contributes to weaker longitudinal and transverse magnetic-field gradients. The weaker longitudinal gradient suppresses the conversion of charge noise into qubit-frequency fluctuations, whereas the weaker transverse gradient reduces electrically driven spin-resonance efficiency and makes drive-dependent errors relevant at lower Rabi frequencies. Additional BEOL routing and metal levels required for two-dimensional arrays make such placement decisions increasingly consequential when scaling. These results establish an integration-level control--noise trade-off linking micromagnet placement within densely routed layouts to achievable gate performance, while very high-fidelity operation remains possible in industrially fabricated devices.
\end{abstract}


\clearpage
\section{Introduction}\label{sec:introduction}
Fault-tolerant quantum computing is a central goal of the second quantum revolution \cite{Dowling2003-sx}, promising transformative applications in quantum chemistry \cite{Abrams1999-lc}, materials discovery \cite{Berry2019-md, Feynman1982-wh}, and large-scale optimization \cite{Farhi2022-uv}. Because quantum information is intrinsically fragile and susceptible to environmental noise, large-scale computation requires quantum error correction to stabilize logical states over long algorithms \cite{fowler_surface_2012}. The substantial overhead associated with error correction implies that practical quantum processors must support millions of physical qubits, placing stringent demands on device performance, uniformity, and scalability \cite{DiVincenzo2000, fowler_surface_2012, Gidney2021-lk}.

Among the various platforms under development \cite{Ladd2010-tf,deLeon2021}, spin qubits in silicon \cite{loss_quantum_1998} offer several key advantages for large-scale quantum computing. They combine long coherence times, operation fidelities exceeding fault-tolerance thresholds \cite{yoneda_quantum-dot_2018,noiri_fast_2022,xue_quantum_2022,Takeda2024-fm}, and compact footprints with compatibility with advanced semiconductor manufacturing \cite{zwerver_qubits_2022,camenzind_spin_2022,Steinacker2025,neyens_probing_2024,hrl_2026}. This compatibility provides a clear pathway toward rapid scaling once a suitable architecture is established. In addition, intrinsically low power consumption \cite{Vandersypen_interfacing_2017,veldhorst_silicon_2017}, operation at elevated temperatures \cite{yang_operation_2020,camenzind_spin_2022}, and the potential for dense integration \cite{Li2025-yd,Vandersypen_interfacing_2017,boter_spiderweb_2022,ha_flexible_2022}, together with the integration of control electronics \cite{petit_universal_2020,xue_cmos-based_2021,Bartee2025,hrl_2026}, further strengthen the prospects of silicon spin qubits for realizing large-scale fault-tolerant quantum computing.

Despite the compatibility with advanced semiconductor manufacturing, progress in silicon spin qubits has so far been driven primarily by academic research. These efforts have established high-fidelity single-qubit \cite{Wu2025-highfidelity} and two-qubit gates \cite{noiri_fast_2022, xue_quantum_2022, mills_two-qubit_2022}, high-fidelity state preparation and measurement \cite{Takeda2024-fm,Huang2024}, and initial demonstrations of quantum error correction \cite{takeda_quantum_2022, hrl_2026}. In parallel, industrial fabrication of silicon spin qubits using 300 mm CMOS lines has recently produced operational qubit devices with characterized coherence and gate performance \cite{neyens_probing_2024, George2025-12qarrays}. This represents a qualitative step toward the process control, device uniformity, and throughput essential for scaling, and makes it possible to assess how dense device integration affects qubit operation.

Here, we study silicon spin qubits fabricated using Intel's 300~mm process and hosted in isotopically enriched ${}^{28}\mathrm{Si}/\mathrm{SiGe}$ heterostructures. A high-performance RIKEN device provides a complementary reference against which the effects of the present Intel micromagnet integration on charge-noise susceptibility and EDSR efficiency can be assessed. Although the heterostructures are nominally similar and isotopically enriched to the same residual $^{29}$Si concentration, the devices differ both in fabrication process and in the integration of qubits, micromagnets, and control wiring. The Intel micromagnets are integrated inline with the back-end-of-line (BEOL) contact routing; their larger separation from the quantum well compared to the RIKEN reference is set by the implemented vertical stack and contributes to weaker longitudinal and transverse magnetic-field gradients. Despite the reduced gradients resulting from the present micromagnet integration, primitive $\pi/2$-gate fidelities reach the 99.99\% level. The measured Intel qubits additionally show reduced qubit-to-qubit variability in microwave-control characteristics, including the operating microwave power and several driven-control quality factors, consistent with prior wafer-scale evidence for tighter process control in advanced 300~mm manufacturing~\cite{neyens_probing_2024}. Together, coherence, noise-correlation and randomized-benchmarking measurements identify an integration-level control--noise trade-off: the weaker longitudinal gradient reduces charge-noise conversion and extends idle coherence, whereas the weaker transverse gradient lowers EDSR efficiency and makes drive-dependent errors relevant at lower Rabi frequencies, limiting the achievable gate fidelity.

\section{Device Integration}\label{sec:results}
\begin{figure}[htbp]
\centering
\includegraphics[width=0.95\linewidth]{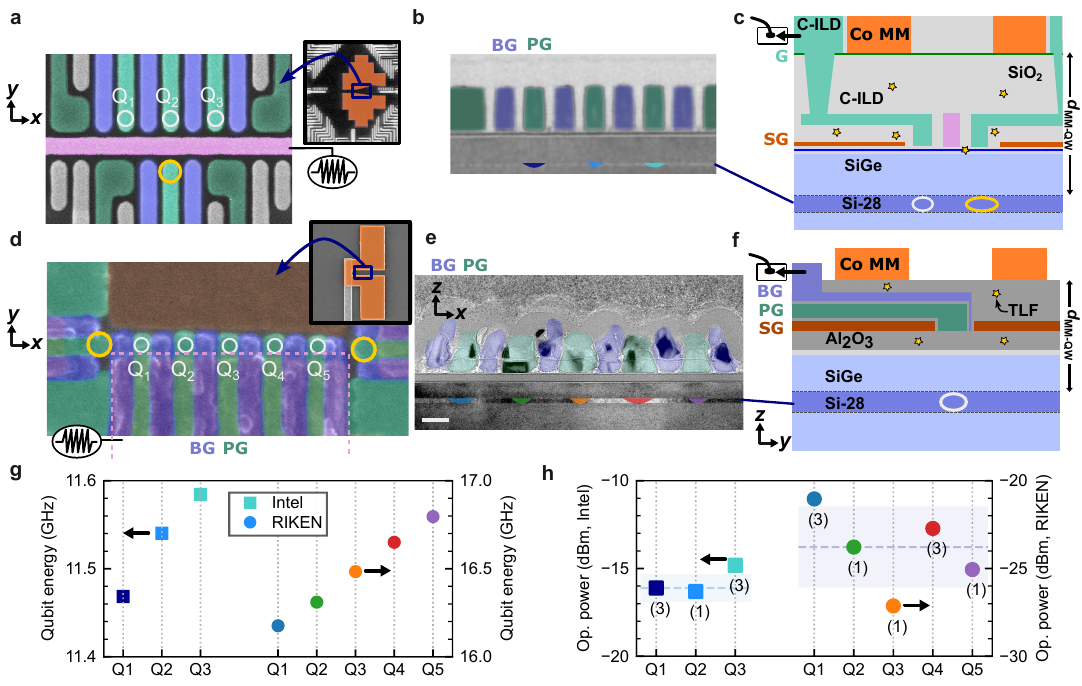}
\caption{\textbf{Industrial and academic silicon spin-qubit devices.}
(a) Scanning electron micrograph of the three-qubit Intel device fabricated in a 300~mm CMOS process using a TiN screening layer and a replacement metal gate (RMG) stack.
(d) Scanning electron micrograph of the five-qubit RIKEN device, showing the overlapping aluminium gate architecture. The colour code identifies barrier gates (BG, purple) and plunger gates (PG, green). Inset: top-down scanning electron micrograph showing the design of the micromagnet fabricated on top of the device (orange).
(b,e) Representative cross-sectional transmission electron micrographs of the industrial (b) and academic (e) devices, showing the gate-electrode geometries and layer stacks.
(c,f) Schematic cross-sections of the gate stacks for the industrial (c) and academic (f) devices. The schematics are not drawn to scale, and the interconnect routing is schematic. Symbols at the upper left indicate the routing path from the gates to bond pads and control electronics: planar fan-out in the academic device and BEOL contact-interconnect routing in the industrial device. The cobalt micromagnet in the industrial device is integrated within the contact interlayer dielectric alongside the routing. Arrows labeled $d_{\mathrm{MM-QW}}$ indicate the micromagnet-quantum-well separation. Screening gates (SG) prevent unwanted accumulation. In the academic device, barrier gates (BG) and plunger gates (PG) are implemented in an overlapping three-layer architecture separated by aluminium oxide (Al$_2$O$_3$), whereas in the industrial device, they are defined within the same gate layer and separated by silicon oxide together with a thin hafnium oxide layer. Yellow stars indicate representative two-level fluctuators (TLFs) in the oxides that give rise to charge noise. The SiGe spacer thickness is 30~nm in the academic device and 50~nm in the industrial device. The larger $d_{\mathrm{MM-QW}}$ in the industrial layout reduces the magnetic-field gradients at the qubit plane.
(g,h) Qubit energies and source microwave powers required to operate the qubits at a Rabi frequency of 1~MHz. The smaller inter-qubit frequency separations and higher median operating power of the industrial qubits are consistent with weaker effective micromagnet gradients, which are set by micromagnet placement, geometry, magnetization, and the external magnetic-field configuration. The narrower operating-power distribution in Intel Device 1 indicates a more reproducible qubit-to-qubit EDSR response across the measured array. The power values are nominal source-calibrated quantities as described in Methods. Numbers in brackets indicate the electron occupation of each quantum dot. Dotted lines and shaded regions correspond to the median and standard deviation of the operating-power distribution for the two device types. The median source microwave powers (± standard deviation across qubits) are $-16.11 \pm 0.81$~dBm for the Intel array and $-23.78 \pm 2.31$~dBm for the RIKEN device. 
}
\label{fig:fig1}
\end{figure}

Figure~\ref{fig:fig1} shows the Intel devices studied here together with the RIKEN reference: three-qubit devices fabricated using Intel's industry-compatible 300\,mm process (Fig.~\ref{fig:fig1}a)~\cite{neyens_probing_2024, George2025-12qarrays} and a five-qubit device fabricated at RIKEN using academic processes (Fig.~\ref{fig:fig1}d)~\cite{Wu2025-highfidelity}. Despite differing fabrication contexts, both types are built on nominally similar heterostructures grown on 300~mm wafers at Intel and 100~mm wafers at QuTech Delft~\cite{PaqueletWuetz2022,DegliEsposti2024}, respectively: isotopically enriched \textsuperscript{28}Si/SiGe quantum wells grown by reduced-pressure chemical vapor deposition using \textsuperscript{28}SiH\textsubscript{4}. The silicon quantum wells hosting the qubits have residual \textsuperscript{29}Si concentrations of approximately 800~ppm.

Structural distinctions relevant to noise nevertheless remain. The SiGe spacer thickness is 30~nm in the academic device and 50~nm in the industrial devices, placing the quantum well closer to the oxide interfaces in the former. Both platforms support electrostatic control of quantum-dot formation and tunnel coupling, including exponential exchange tuning with barrier-gate voltage (Extended Data Fig.~\ref{fig:Fig_Ext_1_exchange}); the greater sensitivity of exchange to barrier-gate voltage in the RIKEN device is consistent with stronger electrostatic coupling from the shallower quantum well, although it is not determined by spacer thickness alone. The closer proximity also increases sensitivity to atomic-scale defects in the device oxides that act as two-level fluctuators, the dominant source of charge noise in spin qubits hosted in purified silicon~\cite{yoneda_quantum-dot_2018, yoneda_noise-correlation_2023, RojasArias2026Scaling, Rojas_Arias_2025_singleshotcorrelations}. The conversion of these charge fluctuations into qubit-frequency noise additionally depends on the longitudinal micromagnet gradient. The quantum-well thickness also differs, being 9~nm in the academic device and approximately 5~nm in the industrial devices. The thinner industrial quantum well can increase wavefunction overlap with the SiGe barriers and may therefore make nuclear spins outside the enriched quantum well, including naturally abundant \textsuperscript{73}Ge, more relevant.

The Intel devices, shown in Fig.~\ref{fig:fig1}a, are realized using an advanced 300~mm fabrication process conducted entirely within the cleanroom and designed to support dense BEOL interconnect routing. They employ a two-layer gate stack comprising a TiN screening layer and a replacement metal gate (RMG) stack, without the overlapping fine-gate architecture used in the academic device (Fig.~\ref{fig:fig1}b). As part of the BEOL integration, a single contact-interconnect layer comprising Cu trench and via structures in a SiO$_2$ contact interlayer dielectric (C-ILD) routes connections between the bonding pads, gate lines, and ohmic contacts~\cite{George2025-12qarrays}. Extending this architecture to quasi-two-dimensional $M \times N$ layouts requires additional routing and metal levels as the number of qubit rows $M$ increases. Micromagnet placement must therefore be designed together with the expanded interconnect stack. Adjacent charge sensors can be integrated alongside the qubit array for charge sensing in one-dimensional or quasi-two-dimensional layouts~\cite{neyens_probing_2024,Li2025-yd}.

The RIKEN device exemplifies academic fabrication, produced in-house in the RIKEN research cleanroom using electron-beam lithography and a flexible process flow.
Figure~\ref{fig:fig1}d shows a scanning electron micrograph of the resulting five-qubit device, which employs three overlapping aluminium gate layers separated by native aluminium oxide (Al$_2$O$_3$); a representative TEM cross-section is shown in Fig.~\ref{fig:fig1}e. In this academic implementation, fan-out to connect the fine-gate features to the bonding pads is achieved using a planar metallization layout. A cobalt micromagnet \cite{yoneda_robust_2015}, shown in the inset of Fig.~\ref{fig:fig1}d, is deposited above the gate stack and electrically isolated from the underlying gates by an atomic-layer-deposited oxide \cite{kawakami_electrical_2014, Pioro-Ladriere_2008}, as illustrated in Fig.~\ref{fig:fig1}f.

In the industrial devices studied here, the cobalt micromagnets are fabricated inline within the contact interlayer dielectric alongside the routing. The micromagnet--quantum-well separation is denoted by $d_{\mathrm{MM-QW}}$ in Fig.~\ref{fig:fig1}c,f. This separation is set by the vertical stack design. The separation is larger than in the academic implementation and, together with the micromagnet geometry, contributes to weaker gradients at the qubit plane. The 20~nm difference in SiGe spacer thickness alone does not determine the separation or the gradient difference. More details on the fabrication of these devices are given in Refs.~\cite{neyens_probing_2024, George2025-12qarrays}.

The Intel devices also show a narrower spread in several measured control-related operating parameters.
Here, the clearest direct operating-control signature is the narrower spread of source microwave powers required for qubit operation (Fig.~\ref{fig:fig1}h), while the driven-control quality factors in Extended Data Fig.~\ref{fig:Fig_Ext_DrivenCoherences} provide additional support. Prior wafer-scale measurements independently show reduced variation in quantum-dot operating voltages across the Intel 300~mm process~\cite{neyens_probing_2024}. The microwave powers reported here are nominal, source-calibrated room-temperature powers rather than direct measurements of the power delivered to the sample (see Methods for details). Such reproducibility can reduce calibration overhead and facilitate shared control in larger qubit arrays~\cite{Vandersypen_interfacing_2017, veldhorst_silicon_2017}.

The micromagnet gradients are central to qubit operation in these devices and serve complementary roles: the longitudinal component $\nabla B_{\parallel}$ creates position-dependent Zeeman shifts for frequency-selective qubit addressability, whereas the transverse component $\nabla B_{\perp}$ couples electrically driven electron motion to the spin, enabling EDSR. The smaller qubit-frequency spacings in the Intel devices ($\sim55~\mathrm{MHz}$ compared with $\sim150~\mathrm{MHz}$ in the RIKEN device; Fig.~\ref{fig:fig1}g) therefore reflect a weaker $\nabla B_{\parallel}$ along the qubit array, consistent with the larger micromagnet--quantum-well separation in the Intel integration. The measured median source microwave power required to reach 1~MHz is $7.7~\mathrm{dB}$ higher in the Intel array than in the RIKEN reference (Fig.~\ref{fig:fig1}h). Simulations yield representative transverse gradients of $\nabla B_{\perp} \approx 1.15~\mathrm{mT/nm}$ for the RIKEN device and $\nabla B_{\perp} = 0.4\text{-}0.5~\mathrm{mT/nm}$ for the Intel device~\cite{neyens_probing_2024}. For micromagnet-based EDSR, the Rabi frequency scales as $f_{\mathrm{R}} \propto \nabla B_{\perp} E_{\mathrm{MW}}$, such that a reduction in transverse gradient requires a higher microwave drive amplitude $E_{\mathrm{MW}}$ and hence higher microwave power \cite{Pioro-Ladriere_2008}. The measured power increase falls within the range predicted from these gradients for comparable dot confinement and gate coupling (see Methods). Thus, the Intel micromagnet integration creates a control--noise trade-off: the weaker longitudinal gradient reduces charge-noise susceptibility, while the weaker transverse gradient requires stronger microwave drive, causing drive-dependent errors to limit the achievable fidelity.

\section{Idling Coherence}
Coherence measurements probe how the magnetic-field gradients and device characteristics established in Fig.~\ref{fig:fig1} affect qubit dephasing. We measure Ramsey $T_{2}^{*}$, Hahn-echo, and dynamical-decoupling coherence times, which probe complementary frequency ranges of the noise spectrum.

\begin{figure}[htbp]
\centering
\includegraphics[width=0.95\linewidth]{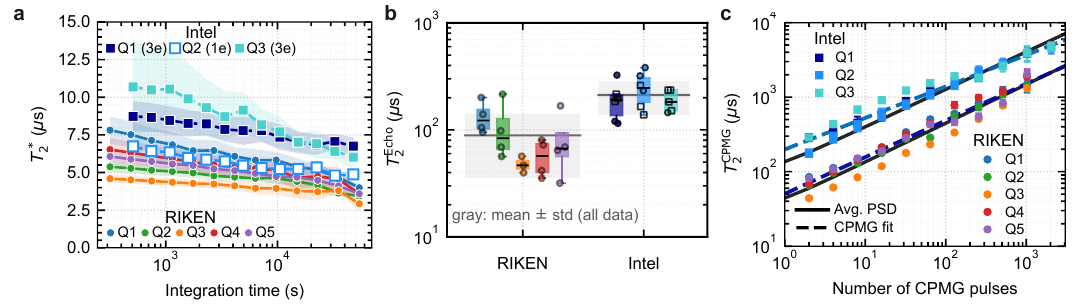}
\caption{\textbf{Idling coherence in academic and industrial devices.}
(a) Ramsey dephasing time $T_{2}^{*}$ as a function of integration time, extracted from independent fits to non-overlapping time segments of continuous Ramsey data. Markers show the median fitted $T_{2}^{*}$, and shaded bands indicate the standard deviation across segments. Qubits in industrial devices exhibit longer $T_{2}^{*}$, consistent with reduced sensitivity to electric-field noise due to a weaker longitudinal magnetic-field gradient $\nabla B_{\parallel}$. Within the industrial platform, qubits in one- and three-electron configurations exhibit distinct $T_{2}^{*}$ values. The reduction of $T_{2}^{*}$ at long integration times in the academic device is attributed to magnetic-field drifts of order 10~Hz/s (see Methods) \cite{RojasArias2026Scaling}.
(b) Hahn-echo coherence times $T_{2}^{\mathrm{echo}}$ measured across several bias configurations, as described in Methods.
Echo decoupling enhances coherence and strongly suppresses the occupation-dependent contrast between one- and three-electron quantum dots, consistent with filtering of slow nuclear-spin-related fluctuations. 
For the industrial platform, circles indicate data from Device 1, and rectangles indicate data from Device 2.
The grey line and shaded region indicate the mean and standard deviation across the plotted qubit/configuration measurements. The colour code matches panel (a).
(c) CPMG coherence times as a function of the number of decoupling pulses for the academic device and one industrial device (Device~2). Dashed curves show fits of the form $T_{2}^{\mathrm{CPMG}}=A\cdot n^{\beta}$ to the corresponding datasets, while the solid black curves indicate coherence times expected from the representative platform noise power spectral densities discussed in Fig.~\ref{fig:fig3}~\cite{yoneda_quantum-dot_2018}.
} 
\label{fig:fig2}
\end{figure}

Figure~\ref{fig:fig2}a shows the Ramsey dephasing time $T_{2}^{*}$ extracted from segmented Ramsey measurements as a function of integration time~\cite{yoneda_noise-correlation_2023,nakajima_coherence_2020}. In a Ramsey measurement, $T_{2}^{*}$ reflects the accumulated effect of qubit-frequency fluctuations during the free-evolution time and is in practice dominated by the lowest-frequency components sampled during the measurement. For both platforms, $T_{2}^{*}$ decreases with increasing integration time, consistent with the growing sensitivity to progressively slower frequency fluctuations. The Intel qubits generally exhibit longer $T_{2}^{*}$ than the RIKEN qubits, particularly at longer integration times where slower fluctuations contribute more strongly, consistent with reduced sensitivity to low-frequency electric-field noise due to the weaker longitudinal magnetic-field gradient $\nabla B_{\parallel}$. Notably, within the industrial platform, $T_{2}^{*}$ separates into two distinct groups corresponding to qubits hosted in one- and three-electron quantum dots. This occupation dependence points to an additional low-frequency contribution discussed in the next section. No comparable separation is observed in the academic device, where stronger coupling to charge noise dominates the low-frequency dephasing.

To suppress quasi-static and low-frequency fluctuations, we perform Hahn-echo measurements (Fig.~\ref{fig:fig2}b). Echo decoupling enhances coherence for both platforms and strongly suppresses the occupation-dependent separation observed in Ramsey $T_{2}^{*}$, indicating that this contrast originates predominantly from slow noise, consistent with the low-frequency nuclear-spin contribution discussed below. Under Hahn echo, the Intel (RIKEN) qubits exhibit median coherence times $T_{2}^{\mathrm{echo}} \approx 200~\mu\mathrm{s}$ ($\approx 70~\mu\mathrm{s}$), consistent with reduced charge-noise susceptibility arising from the weaker longitudinal magnetic-field gradient in the industrial devices.
As described in Methods, the Hahn-echo data in Fig.~\ref{fig:fig2}b were acquired over several bias configurations within the operating window of each device.

Under Carr-Purcell-Meiboom-Gill (CPMG) dynamical decoupling, coherence times increase systematically with the number of refocusing pulses $n$ for both platforms. Consistent with the trends observed in $T_{2}^{*}$ and $T_{2}^{\mathrm{echo}}$, the industrial device exhibits longer coherence times across all pulse numbers. This difference is evident at $n = 1024$, where the two Intel qubits reach a median coherence time of $T_{2}^{\mathrm{CPMG}} = 3.88 \pm 0.09~\mathrm{ms}$, compared with $1.49 \pm 0.31~\mathrm{ms}$ across five RIKEN qubits, with uncertainties denoting the standard deviation across qubits.

To assess the scaling with pulse number, we fit the median coherence times to a power-law dependence, $T_{2}^{\mathrm{CPMG}} = A\cdot n^{\beta}$. This yields scaling exponents $\beta = 0.422 \pm 0.010$ and $\beta = 0.492 \pm 0.007$ for the Intel and RIKEN datasets, respectively. The corresponding prefactors $A$ are in agreement with the independently extracted Hahn-echo coherence times (Fig.~\ref{fig:fig2}b). Although the coherence times differ substantially, the scaling exponents are broadly consistent with the $\beta = 0.5$ behavior expected for decoherence dominated by $1/f$-like noise, commonly associated with charge noise arising from ensembles of two-level fluctuators \cite{yoneda_quantum-dot_2018, yoneda_noise-correlation_2023, RojasArias2026Scaling, Rojas_Arias_2025_singleshotcorrelations}. This indicates similar $1/f$-like high-frequency spectral scaling in the two platforms despite their different noise amplitudes, consistent with the noise spectra analyzed below (Fig.~\ref{fig:fig3}a).

\section{Noise Characterization and Correlations}
\begin{figure}[htbp]
\centering
\includegraphics[width=0.95\linewidth]{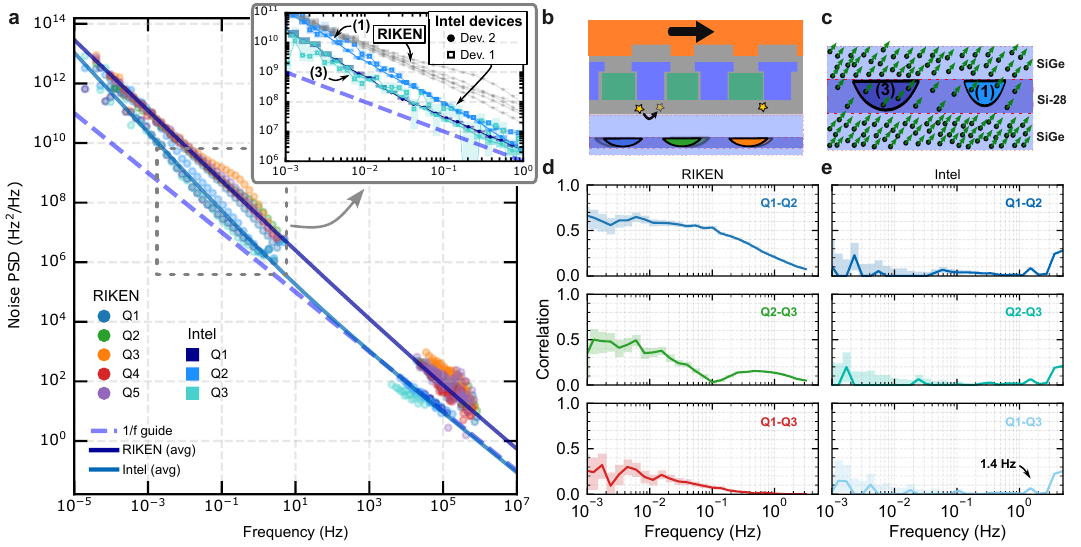}
\caption{ \textbf{Noise spectra and noise correlations.}
(a) Power spectral densities (PSDs) spanning the low-frequency and CPMG regimes for RIKEN and Intel qubits. Solid curves show representative PSDs constructed from the median fit parameters across qubits of each platform (see Methods). The purple dashed line shows a $1/f$ continuation of the high-frequency charge-noise contribution in the industrial device, highlighting excess low-frequency noise beyond the extrapolated charge-noise level.
Inset: Enlarged view of the low-frequency regime. RIKEN qubits exhibit predominantly charge-noise-dominated behavior, whereas the industrial platform shows a pronounced low-frequency contribution consistent with residual nuclear-spin noise. The three-electron operating configurations exhibit lower low-frequency noise than the one-electron configurations.
(b) Schematic illustration of correlated charge noise: a switching two-level fluctuator (TLF) capacitively couples to multiple qubits. TLF switching induces correlated shifts in the electron positions, which translate into correlated qubit-frequency fluctuations via the longitudinal magnetic-field gradients.
(c) Illustration of nuclear-spin-carrying isotopes in the Si/SiGe heterostructure (natural abundances: 4.7\% $^{29}$Si and 7.75\% $^{73}$Ge). The isotopically enriched silicon quantum well contains 0.08\% $^{29}$Si. The few nuclei shown are schematic; at this enrichment level, a typical qubit wavefunction overlaps with on the order of $10^{2}$ residual $^{29}$Si nuclei in the quantum well. 
(d) Noise-correlation magnitudes for RIKEN qubit pairs, showing pronounced low-frequency Q1-Q2 correlations and weaker but finite next-nearest-neighbour Q1-Q3 correlations. The minimum in the Q2-Q3 correlation magnitude near 0.1 Hz coincides with a sign reversal of the real part of the cross-spectrum \cite{RojasArias2026Scaling, yoneda_noise-correlation_2023}, consistent with a crossover between contributions from different two-level fluctuators (TLFs).
(e) Corresponding data for Intel Device~1, showing much weaker correlations, consistent with a larger contribution of uncorrelated nuclear-spin noise. Narrow correlation peaks arise from global electrical noise, including a feature consistent with the refrigerator pulse tube ($\sim$1.4~Hz). Measurements on Device~2 showed similar inter-qubit correlation magnitudes over the measured frequency band.
}
\label{fig:fig3}
\end{figure}

To explain the coherence trends in Fig.~\ref{fig:fig2}, we analyze qubit-frequency fluctuations extracted from continuous Ramsey-frequency tracking~\cite{RojasArias2026Scaling, Rojas_Arias_2025_singleshotcorrelations, yoneda_noise-correlation_2023} and CPMG measurements~\cite{yoneda_quantum-dot_2018}. In silicon spin qubits, charge noise originates from two-level fluctuators (TLFs) in the surrounding oxides, including atomic-scale defects such as tunneling atoms or dangling bonds, which generate local electric-field fluctuations. Figure~\ref{fig:fig3}a shows representative power spectral densities (PSDs) derived from the measured frequency time traces.

At the higher frequencies accessed via the CPMG protocol~\cite{yoneda_quantum-dot_2018}, both platforms exhibit a $1/f$-like spectrum characteristic of charge noise, but the qubit-frequency-noise level is lower in the industrial devices. This reduction is consistent with the weaker longitudinal magnetic-field gradient $\nabla B_{\parallel}$ inferred from the qubit-frequency spacings in Fig.~\ref{fig:fig1}g. Charge-noise-induced electric-field fluctuations are converted into qubit-frequency fluctuations through $\nabla B_{\parallel}$, so a weaker gradient reduces the qubit-frequency response to a given charge-noise environment. This interpretation is supported by the independently measured sensor-referred charge-noise spectra in Extended Data Fig.~\ref{fig:Fig_Ext_2_sensornoise}, which show comparable local charge-noise levels in the two platforms. The reduced qubit-frequency noise therefore reflects reduced susceptibility to charge noise, rather than implying a lower underlying charge-noise amplitude in the industrial devices.

At low frequencies, both platforms deviate from a pure $1/f$ charge-noise spectrum, but for distinct physical reasons, as illustrated in Fig.~\ref{fig:fig3}b,c. In the academic device, pronounced qubit-to-qubit variations and Lorentzian-like features from individual nearby TLFs are observed on top of the overall $1/f$ background. The visibility of TLFs with switching frequencies around 1~Hz is consistent with the larger longitudinal magnetic-field gradients and stronger coupling to the oxide environment in the shallower quantum well. The qubit-to-qubit variability further indicates that different qubits couple to distinct local TLF environments.

The low-frequency deviation from $1/f$ behavior is markedly stronger in the industrial devices, where the measured noise exceeds the extrapolated charge-noise contribution (inset in Fig.~\ref{fig:fig3}a). We attribute this excess low-frequency noise to residual nuclear spins, including residual $^{29}$Si at approximately 800~ppm in the quantum well. This interpretation is further supported by the separation between one- and three-electron operating configurations observed in $T_{2}^{*}$ (Fig.~\ref{fig:fig2}a), the similar low-frequency noise of the three-electron qubits, and the reversible change in noise when the electron occupations are exchanged. The lower noise in the three-electron configurations reflects weaker effective confinement associated with the higher accumulation voltages used for these configurations. The resulting more extended electronic wavefunction increases spatial averaging of the nuclear Overhauser field and reduces its contribution to qubit-frequency noise~\cite{Malinowski2017,Struck2020}.

Beyond the noise amplitude, a key qualitative distinction between the platforms emerges from inter-qubit correlations. Electric-field fluctuations from a shared TLF can capacitively couple to multiple qubits, leading to correlated or anti-correlated qubit-frequency shifts depending on the relative positions of the qubits with respect to the fluctuator, as illustrated in Fig.~\ref{fig:fig3}b. Consistent with this picture, Fig.~\ref{fig:fig3}d shows pronounced noise correlations between adjacent RIKEN qubits. These correlations generally decrease with qubit separation, with a frequency-dependent relative magnitude, consistent with individual TLFs influencing multiple nearby qubits over a finite spatial range \cite{RojasArias2026Scaling}.

By contrast, much weaker correlations are observed in the industrial device (Fig.~\ref{fig:fig3}e). This behavior is consistent with the uncorrelated nature of nuclear-spin noise at low frequencies, where Overhauser-field fluctuations arise from independent ensembles of nuclear spins localized within each quantum dot. The reduced correlation magnitudes therefore support the interpretation that uncorrelated residual nuclear-spin noise dominates the low-frequency spectrum, masking weaker charge-noise correlations. Further isotopic purification would reduce this nuclear-spin contribution and could therefore make charge noise and its associated correlations relatively more prominent at low frequencies. At higher frequencies, where charge noise dominates, correlations may become more relevant.

\section{Single-Qubit Gate Performance}
\begin{figure}[htbp]
\centering
\includegraphics[width=0.95\linewidth]{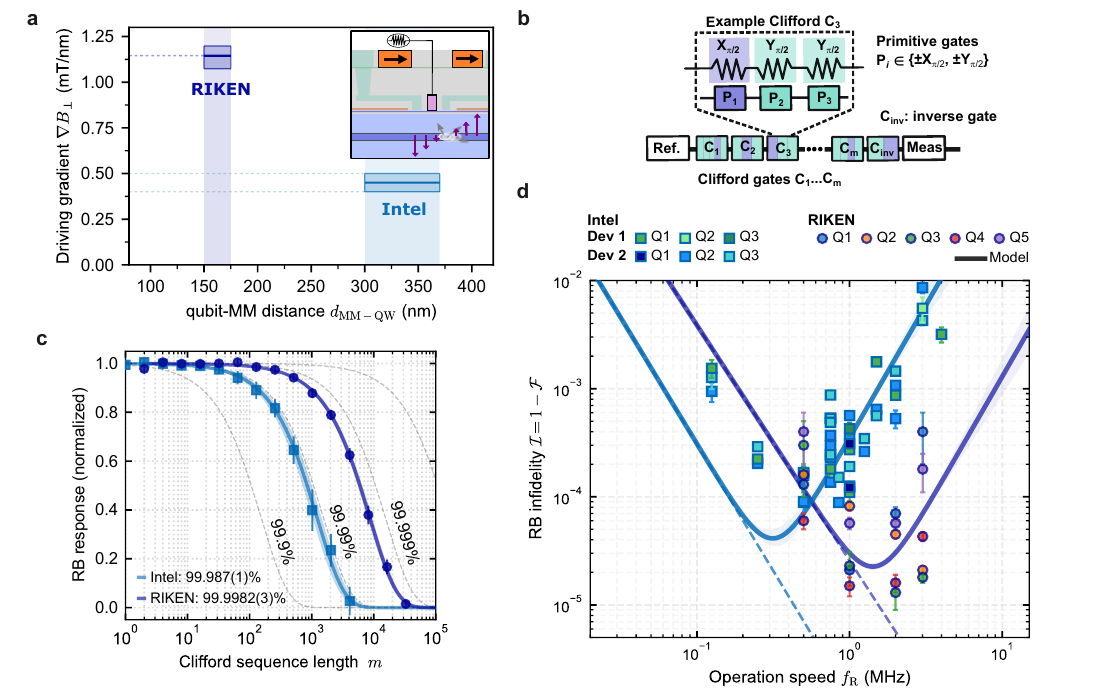}
\caption{\textbf{Single-qubit gate performance and integration-dependent drive limitations.}
(a) Simulated transverse driving gradients $\nabla B_{\perp}$ for representative Intel and RIKEN micromagnet geometries at their respective approximate micromagnet--quantum-well separations $d_{\mathrm{MM-QW}}$. The coloured bars show representative gradients from magnetostatic simulations of 0.4-0.5~mT/nm (Intel) and approximately 1.15~mT/nm (RIKEN). Horizontal bar widths and shaded bands indicate approximate separation ranges. The Intel (RIKEN) calculations are evaluated at an external magnetic field of 0.35~T (0.45~T). The inset schematically illustrates the micromagnet gate stack and the resulting transverse magnetic field $B_{\perp}$ at the qubit plane.
(b) Randomized benchmarking protocol. Following initialization, $m$ random single-qubit Cliffords $C_i$ are applied, followed by a recovery Clifford $C_{\mathrm{inv}}$ that ideally returns the qubit to the target state before measurement. All Cliffords are compiled exclusively from primitive rotations $P_i\in\{\pm X/2,\pm Y/2\}$.
(c) Representative high-fidelity RB decay curves for Intel and RIKEN qubits. The normalized RB response is plotted as a function of the Clifford sequence length $m$ and fitted to extract primitive $\pi/2$-gate fidelities of $\sim99.998\%$ and $\sim99.99\%$ for the RIKEN and Intel qubits, respectively. Dashed curves indicate reference decays corresponding to primitive $\pi/2$-gate fidelities from $99.9\%$ to $99.9999\%$.
(d) RB primitive $\pi/2$-gate infidelity $\mathcal{I}$ as a function of Rabi frequency $f_{\mathrm{R}}$ for all measured qubits. Dashed curves show the phase-noise limit $\mathcal{I}_{\mathrm{ph}}(f_{\mathrm{R}})$ obtained from the measured noise power spectral densities. Solid curves show $\mathcal{I}(f_{\mathrm{R}})=\mathcal{I}_{\mathrm{ph}}(f_{\mathrm{R}})+a_{\mathrm{d}}f_{\mathrm{R}}^{n_{\mathrm{RB}}}$, with $n_{\mathrm{RB}}=2.46$ obtained from the Intel RB-infidelity fit and $a_{\mathrm{d}}$ fitted separately for the two platforms. Shaded bands indicate the central 68\% interval obtained by propagating the uncertainty in $n_{\mathrm{RB}}$. The RIKEN device remains close to the phase-noise limit up to higher operation speeds, whereas the Intel devices deviate at lower $f_{\mathrm{R}}$, consistent with their larger microwave-power requirement for control.  Intel error bars indicate $\pm1$ bootstrap standard deviation, obtained by resampling complete acquisition rounds within each RB measurement and refitting the averaged decay for 300 resamples. More details on the RB experiments are given in Methods.
}
\label{fig:fig4}
\end{figure}

We next examine how the weaker transverse magnetic-field gradient in the Intel devices affects single-qubit control. Figure~\ref{fig:fig4}a compares representative simulated driving gradients $\nabla B_{\perp}$ for the RIKEN and Intel micromagnet geometries at their respective approximate micromagnet--quantum-well separations $d_{\mathrm{MM-QW}}$. The larger $d_{\mathrm{MM-QW}}$ in the Intel devices yields a substantially weaker $\nabla B_{\perp}$, consistent with the higher source microwave power required to reach a given Rabi frequency in Fig.~\ref{fig:fig1}h.

We characterize single-qubit gate performance using randomized benchmarking (RB), illustrated in Fig.~\ref{fig:fig4}b, with representative decay curves shown in Fig.~\ref{fig:fig4}c and the extracted RB primitive $\pi/2$-gate infidelity $\mathcal{I} \equiv 1-\mathcal{F}$ plotted versus $f_{\mathrm{R}}$ in Fig.~\ref{fig:fig4}d. Despite the weaker control gradient, the Intel devices achieve primitive $\pi/2$-gate fidelities near $99.99\%$, while the RIKEN reference reaches $99.999\%$. The higher drive amplitude required in the Intel devices is therefore expected to make drive-dependent errors relevant at lower $f_{\mathrm{R}}$.

At low to moderate operation speeds, the RB infidelity decreases with increasing $f_{\mathrm{R}}$, consistent with a decreasing contribution from phase noise. In this regime, qubit-frequency fluctuations accumulated during the finite gate time contribute to the gate error, such that faster gates are exposed to low-frequency noise for a shorter duration. The dashed curves in Fig.~\ref{fig:fig4}d show the corresponding phase-noise limit $\mathcal{I}_{\mathrm{ph}}(f_{\mathrm{R}})$ obtained from simulations based on the measured noise power spectral densities (PSDs) in Fig.~\ref{fig:fig3}a.

At larger $f_{\mathrm{R}}$, both platforms deviate upward from $\mathcal{I}_{\mathrm{ph}}(f_{\mathrm{R}})$, indicating an additional drive-dependent contribution to the gate infidelity. We describe the RB data using $\mathcal{I}(f_{\mathrm{R}})=\mathcal{I}_{\mathrm{ph}}(f_{\mathrm{R}})+a_{\mathrm{d}}f_{\mathrm{R}}^{n_{\mathrm{RB}}}$, shown by the solid curves in Fig.~\ref{fig:fig4}d. Fitting the Intel data gives $n_{\mathrm{RB}}=2.46\pm0.28$, which we use for both platforms. The independently measured increase in spin-lock relaxation rates $1/T_{1\rho}$ with microwave power further supports a drive-dependent decoherence contribution (Extended Data Fig.~\ref{fig:Fig_Ext_DrivenCoherences})~\cite{Takeda2026_RB99999}.

MW-heating measurements reveal qubit-frequency shifts in both platforms, which remain modest and often partially saturate at longer exposure times (Extended Data Fig.~\ref{fig:Fig_Ext_MWHeating})~\cite{undseth_hotter_2023,Wu2025-highfidelity}. Even when summarized by an effective linear detuning rate, the measured shifts are substantially smaller than required to reproduce the high-power infidelities. Thus, the measured microwave-induced detuning shifts cannot account for the observed drive-dependent degradation. Independent off-resonant coherence measurements show little systematic power dependence of Ramsey dephasing, whereas the CPMG-8 decay rate and reconstructed qubit-frequency-noise PSD at $10~\mathrm{kHz}$ increase strongly with microwave power (Extended Data Fig.~\ref{fig:Fig_Ext_CPMG_Ramsey}). Together with the stronger power dependence of the spin-lock decay rate $1/T_{1\rho}$, which probes longitudinal noise near $f_{\mathrm{R}}$~\cite{Takeda2026_RB99999}, these measurements are consistent with a microwave-induced increase in qubit-frequency noise, rather than a simple quasi-static detuning shift, as the relevant high-power limitation. These results show that the weaker $\nabla B_{\perp}$ in the present Intel integration moves the onset of drive-dependent gate errors to lower Rabi frequencies, thereby limiting the peak single-qubit fidelity.

\section{Discussion}\label{sec:discussion}
The Intel devices studied here demonstrate an integration-level control--noise trade-off relevant to scaling. In the Intel devices, the micromagnets are integrated inline with the BEOL contact routing at a larger micromagnet--quantum-well separation $d_{\mathrm{MM-QW}}$ than in the RIKEN reference. The weaker $\nabla B_{\parallel}$ suppresses the conversion of low-frequency electric-field noise into qubit-frequency fluctuations, whereas the weaker $\nabla B_{\perp}$ reduces EDSR efficiency and requires higher drive amplitude to reach the same Rabi frequency. Despite this reduced control efficiency, the Intel devices achieve primitive $\pi/2$-gate fidelities near $99.99\%$, although drive-dependent errors become relevant at lower $f_{\mathrm{R}}$ and limit the achievable fidelity. The RIKEN device provides the complementary strong-gradient reference, with more efficient low-power control but stronger conversion of charge noise into qubit-frequency fluctuations and stronger low-frequency inter-qubit correlations. With further scaling, an integration trade-off becomes increasingly relevant: if additional BEOL interconnect levels and the associated dielectric layers are required while the micromagnet remains in the topmost interconnect level, $d_{\mathrm{MM-QW}}$ increases further and the available magnetic-field gradients weaken. This constraint arises from this placement strategy, rather than industrial fabrication itself, and also applies to academic devices adopting dense multilevel routing.

One possible mitigation path is to place micromagnets in lower metal or C-ILD levels closer to the qubit plane, thereby increasing the available driving gradient $\nabla B_{\perp}$ and reducing the microwave power required for control. Such an approach, however, competes directly with routing and wiring density, since regions occupied by extended micromagnets cannot simultaneously serve as signal interconnects within the same level. An alternative strategy is to use more localized magnetic structures or magnetic elements integrated directly into the gate stack~\cite{Tadokoro_designs_2021,Unseld_baseband_2025,Bersano_magnetintegration_2025,Aldeghi_nanomagnets_2025}. Whether such structures can provide sufficient $\nabla B_{\perp}$ for efficient high-fidelity control while maintaining the required longitudinal field profile in densely integrated arrays remains an open question.

Importantly, the impact of correlated noise must be assessed at the level of quantum error correction rather than individual qubits. Such correlations are not specific to micromagnet devices and can also arise in other electrically sensitive spin-qubit platforms, including hole-spin and exchange-based qubits. Prior work has shown that the degree of qubit-qubit correlations observed in academic micromagnet devices leads to a manageable reduction in error-correction efficiency, corresponding to a modest and scalable qubit overhead rather than a fundamental limitation \cite{RojasArias2026Scaling}. Moreover, architectures using electron shuttling or dynamic reconfiguration can increase qubit spacing, suppressing correlations mediated by shared TLFs while retaining strong local gradients where needed.

\section{Conclusion}\label{sec:conclusion}
From a scalability perspective, the industrial devices studied here demonstrate that ${}^{28}\mathrm{Si}/\mathrm{SiGe}$ spin qubits produced in an industry-compatible 300\,mm process can achieve primitive $\pi/2$-gate fidelities near 99.99\% while exhibiting reduced qubit-to-qubit variability across several measured characteristics. These results build on the process uniformity, reproducibility, and dense integration enabled by industrial fabrication \cite{neyens_probing_2024, George2025-12qarrays}. At the same time, the larger micromagnet--quantum-well separation in the present Intel BEOL integration weakens the magnetic-field gradients, reducing the conversion of charge noise into qubit-frequency fluctuations but also lowering EDSR efficiency and making drive-dependent errors relevant at lower Rabi frequencies. Further scaling of densely routed micromagnet-based spin-qubit processors will therefore require routing architecture and micromagnet placement to be co-designed, maintaining the transverse gradients needed for efficient EDSR while limiting charge-noise sensitivity through the longitudinal gradient.

\clearpage
\section{Methods}\label{sec:methods}

\subsection*{Devices and Experimental Conditions}

\paragraph{Devices.}
The academic device is a gate-defined Si/SiGe spin-qubit array formed in a 9 nm thick isotopically enriched Si quantum well (residual $^{29}$Si concentration $\sim 800$ ppm). The wafer was shipped in air from QuTech Delft to RIKEN before device fabrication. The quantum well is covered by a 30 nm SiGe spacer layer and an ultrathin SiO$_x$ layer (approximately 2 nm). A 15 nm Al$_2$O$_3$ layer is deposited by atomic layer deposition (ALD), followed by three layers of overlapping aluminium gates patterned by electron-beam lithography and lift-off. Inter-layer electrical isolation is provided by native AlO$_x$ formed by controlled oxidation of the aluminium gates. A micromagnet (5 nm Ti / 250 nm Co) is fabricated above the gate stack and separated from it by a 30 nm ALD Al$_2$O$_3$ layer; the design follows prior work, with a characteristic gap size of $\sim 400$ nm \cite{noiri_fast_2022,takeda_fault-tolerant_2016,yoneda_robust_2015}. More details on this and a nominally identical device are found in Ref.~\cite{Wu2025-highfidelity, Takeda2026_RB99999}.

The industrial devices are three-qubit Si/SiGe spin-qubit devices fabricated using Intel's industry-compatible 300\,mm process with integrated micromagnets. The wafers remained within the cleanroom throughout fabrication. For the devices measured here, cobalt micromagnets were fabricated inline within the contact interlayer dielectric using a wafer-level electron-beam process, alongside the interconnect routing. The measurements were acquired on two nominally similar devices, denoted Device~1 and Device~2, with device assignments specified in the corresponding figure captions where relevant. More details on the fabrication of these devices are given in Refs.~\cite{neyens_probing_2024, George2025-12qarrays}.

\paragraph{Device operation.}
Quantum dots, tunnel couplings, and qubit operating points are controlled and tuned using DC gate voltages. To maximize the Pauli spin blockade (PSB) readout window, the academic device is operated in the $(3,1,1,3,1)$ charge configuration, while the industrial devices are operated in the $(3,1,3)$ charge configuration \cite{philips_universal_2022}.  To check the occupation dependence discussed in Figs.~\ref{fig:fig2}a and \ref{fig:fig3}a, we also retuned the industrial device to other accessible charge configurations, including configurations in which the electron occupations were exchanged between dots, such as $(0,3,1)$.

During microwave control, residual coupling to electron reservoirs was minimized to suppress reservoir-assisted charge motion and relaxation. In the industrial devices, this was implemented by ramping down the reservoir gates after initial tuning when this improved gate performance~\cite{Takeda2026_RB99999}. In the academic device, the relevant reservoirs are sensor quantum dots that are energetically detuned from the qubit array during microwave operation.

During these experiments, data were acquired at several bias configurations within the operating window of each device. The screening-gate voltages were adjusted by small amounts, typically tens to hundreds of millivolts, to maintain stable accumulation and sensor operation. These changes were accompanied by compensating adjustments to the plunger and barrier gates to restore the target charge configuration, tunnel couplings, and readout conditions. The spread of Hahn-echo data in Fig.~\ref{fig:fig2}b therefore reflects locally retuned operating points. The randomized-benchmarking data were acquired after manual optimization of the electrostatic operating point and microwave calibration \cite{Wu2025-highfidelity}, and most coherence measurements were performed in the same operating regime in which high single-qubit fidelities, reliable Pauli-spin-blockade readout, and low residual exchange during single-qubit operation were obtained. This context is important because charge-noise sensitivity and, in particular, inter-qubit noise correlations in micromagnet spin qubits can depend strongly on gate voltages, especially barrier-gate settings~\cite{RojasArias2026Scaling}.

\paragraph{Control and pulse sequences.}
Qubit manipulation is performed using electric-dipole spin resonance driven by microwave control pulses applied to gate electrodes in the presence of micromagnet-induced magnetic-field gradients. Baseband voltage pulses are used to move between operation and readout points and to control inter-dot couplings. The room-temperature control hardware, waveform generation, and calibration procedures are similar for the academic and industrial devices and are described in detail in Ref.~\cite{Wu2025-highfidelity}.

In the academic device, microwave control pulses with a Kaiser envelope are used for high-fidelity control~\cite{Wu2025-highfidelity}.  
In the industrial devices, RB measurements were performed using both square and Kaiser-envelope pulses. Previous measurements have shown that rectangular pulses can bias the measured fidelity through off-resonant excitation of the spectator qubit participating in PSB readout, whereas spectrally tailored pulses, such as Kaiser-envelope pulses, suppress this readout-related artifact~\cite{Takeda2026_RB99999}. In the present Intel data, no systematic difference between square- and Kaiser-envelope measurements is resolved over the investigated operating range (Extended Data Fig.~\ref{fig:Fig_Ext_DrivenCoherences}c). The Kaiser-envelope measurements also exhibit increased infidelity at high drive. A contribution of this readout artifact to individual square-pulse measurements nevertheless cannot be excluded.

\paragraph{Readout.}
Qubit readout is performed using spin-to-charge conversion via parity-mode Pauli spin blockade (PSB). In the academic device, the readout architecture enables extraction of three bits of information per measurement cycle, allowing up to three qubits to be targeted within a single cycle. In the industrial devices, readout is performed in either double-dot or triple-dot mode, yielding one bit of information per measurement cycle. While the academic device is initialized using active reset \cite{philips_universal_2022,Wu2024}, the industrial devices are operated in a differential readout mode (parity before operation versus parity after). Since the protocols discussed here rely on detecting spin flips rather than absolute spin states, and no significant relaxation was observed on a 10~ms timescale in nominally identical samples~\cite{Takeda2026_RB99999}, far exceeding the readout durations of approximately 10~$\mu$s in the academic device and 20-40~$\mu$s in the industrial devices, these different readout and initialization schemes are not expected to measurably affect the extracted results, including the frequency-noise spectra obtained from Ramsey and CPMG experiments.

\paragraph{Magnetic-field drifts.}
Slow magnetic-field drifts are present in the experimental setup of the academic device. For noise power spectral density (PSD) extraction, these drifts can be removed in post-processing as described in Ref.~\cite{RojasArias2026Scaling}. For $T_{2}^{*}$ obtained from long integration-time Ramsey measurements in Fig.~\ref{fig:fig2}a, these drifts cannot be fully removed and therefore reduce the extracted $T_{2}^{*}$ at long averaging times. In contrast, coherence measurements probing shorter time scales (e.g., Hahn-echo and dynamical-decoupling measurements in Fig.~\ref{fig:fig2}), as well as gate operations used to extract single-qubit gate fidelities in Fig.~\ref{fig:fig4}, are not expected to be limited by these long-term drifts at the fidelity levels obtained here.
The superconducting magnet used in the cryogenic setup for the industrial-device measurements exhibited no measurable magnetic-field drifts. We emphasize that magnetic-field stability depends on the specific superconducting magnet, rather than on the qubit device itself.

\subsection*{Power spectral density extraction and fitting}
Low-frequency qubit-frequency fluctuations were extracted from continuous Ramsey-frequency tracking following Refs.~\cite{yoneda_noise-correlation_2023,RojasArias2026Scaling}. In each acquisition cycle, a Ramsey sequence with 100 different free-evolution times was measured using a qubit detuning of approximately 1~MHz, and Bayesian estimation was used to extract the instantaneous qubit frequency from the resulting Ramsey fringe. Repeating this procedure over time yields a frequency time trace for each qubit, from which the low-frequency part of the qubit frequency-noise power spectral density (PSD) is obtained. The same continuous Ramsey data were used to extract the integration-time dependence of Ramsey $T_{2}^{*}$ in Fig.~\ref{fig:fig2}a: the spin-readout records were divided into non-overlapping segments of the corresponding duration after discarding the initial settling repetitions, and each segment was fit independently to a Gaussian-decaying Ramsey fringe. The plotted value is the median of the finite $T_{2}^{*}$ values obtained from the segment-wise fits, and the shaded band indicates their standard deviation across segments. Inter-qubit correlation magnitudes in Fig.~\ref{fig:fig3}d,e were obtained from the corresponding qubit-frequency-fluctuation traces using the cross-spectral analysis described in Refs.~\cite{yoneda_noise-correlation_2023,RojasArias2026Scaling}.

The higher-frequency part of the PSD was inferred from CPMG measurements using the standard filter-function approach \cite{yoneda_quantum-dot_2018}. An $n$-pulse CPMG sequence with interpulse spacing $\tau$ is most sensitive to noise near $f \approx 1/(2\tau)$, such that the measured decay of $T_{2}^{\mathrm{CPMG}}$ provides information about the noise amplitude in that frequency range. Combining the repeated-Ramsey and CPMG data therefore gives access to the PSD over a broad frequency range relevant to the coherence and gate-fidelity measurements discussed here.

For each qubit, the resulting frequency-noise PSD was fit to a phenomenological noise-spectrum model of the form
\[
S_f(f) = A f^{\alpha_1} + B f^{-1},
\]
where the first term captures the dominant low-frequency contribution and the second term represents the $1/f$ charge-noise background. Here $S_f(f)$ denotes the qubit frequency-noise PSD in units of $\mathrm{Hz}^2/\mathrm{Hz}$. All frequency-noise PSDs use a two-sided normalization and only positive frequencies are shown. Representative PSDs shown in Fig.~\ref{fig:fig3}a are constructed from the median fit parameters across qubits within each platform. The resulting representative PSDs capture the overall magnitude of the measured Hahn-echo coherence times and the pulse-number dependence of the CPMG coherence times, demonstrating the internal consistency of the noise analysis.

For the academic (RIKEN) device, the median fit parameters are $A = (3.22 \pm 2.1)\times10^{7}$, $\alpha_1 = -1.189 \pm 0.073$, and $B = (3.90 \pm 5.5)\times10^{6}$. For the industrial (Intel) device, the corresponding values are $A = (2.00 \pm 0.54)\times10^{6}$, $\alpha_1 = -1.345 \pm 0.005$, and $B = (8.67 \pm 7.6)\times10^{5}$. Central values denote medians across qubits within each platform, and quoted spreads are MAD-sigma estimates. The large spreads in $A$ and $B$ arise from correlations between fit parameters and the limited frequency range over which the two noise contributions can be cleanly separated. The larger fitted-parameter spread in the academic device is consistent with qubit-dependent coupling to local two-level fluctuators.

\subsection*{Randomized benchmarking}
Single-qubit gate performance was characterized using standard single-qubit randomized benchmarking (RB) \cite{muhonen_quantifying_2015, harper_statistical_2019,Wu2025-highfidelity, Takeda2026_RB99999}. Each RB Clifford is compiled exclusively from primitive $\pm X/2$ and $\pm Y/2$ rotations, such that the extracted primitive-gate fidelity can be directly attributed to the corresponding $\pi/2$ control pulses. For each sequence length, randomized Clifford sequences were applied and the difference between flip and no-flip probabilities was fit with the standard exponential form $P(n)=Ap^n$, from which the depolarizing parameter $p$ was obtained. The Clifford fidelity was calculated as $F_C=(1+p)/2$ and converted into the primitive $\pi/2$-gate fidelity using an average of 3.25 primitive rotations per Clifford, following Refs.~\cite{Wu2025-highfidelity}.

For the control-related quality factors shown in Extended Data Fig.~\ref{fig:Fig_Ext_DrivenCoherences}, $T_{2}^{\mathrm{Rabi}}$ denotes the decay time of resonantly driven Rabi oscillations, while $T_{1\rho}$ is extracted from an exponential fit to the spin-lock decay with fixed offset 0.5. The RB-derived time is defined from the primitive-gate fidelity $\mathcal{F}$ as
\[
T_{2}^{\mathrm{RB}}=-\frac{t_{\mathrm{gate}}}{\ln \mathcal{F}}, \qquad
t_{\mathrm{gate}}=\frac{1}{4f_{\mathrm{R}}}.
\]
For shaped RB pulses, $f_{\mathrm{R}}$ denotes the equivalent operation speed defined by the primitive $\pi/2$-gate duration, rather than the peak instantaneous Rabi frequency \cite{Wu2025-highfidelity}.
The corresponding quality factors are $Q_{\mathrm{Rabi}}=f_{\mathrm{R}}T_{2}^{\mathrm{Rabi}}$, $Q_{1\rho}=f_{\mathrm{R}}T_{1\rho}$, and $Q_{\mathrm{RB}}=f_{\mathrm{R}}T_{2}^{\mathrm{RB}}$.
Shaded bands indicate pointwise 16th--84th bootstrap percentiles of the smoothed local median, obtained by resampling individual quality-factor measurements within local Rabi-frequency windows separately for each platform.

For the rate-domain analysis, measurements at the same Rabi frequency are grouped and represented by their median microwave power and decay rate. The Intel $\Gamma_{\mathrm{RB}}=1/T_{2}^{\mathrm{RB}}$ data are combined into a shared trace across the measured qubits, whereas $\Gamma_{1\rho}=1/T_{1\rho}$ is analyzed separately for Q2 and Q3. The rates are fitted in log space using
\[
\Gamma(P)=
\begin{cases}
c, & P\leq P_0,\\
c(P/P_0)^{\alpha}, & P>P_0,
\end{cases}
\]
where $c$ is the low-power plateau rate, $P_0$ is the transition power, and $\alpha$ is the high-power exponent. For the Intel RB rates, we fix $\alpha_{\mathrm{RB}}=1.73$ from the infidelity fit described below and fit $c$ and $P_0$. The unweighted spin-lock fits yield $\alpha_{1\rho,\mathrm{Q2}}=1.4\pm0.09$ and $\alpha_{1\rho,\mathrm{Q3}}=1.80\pm0.06$, with uncertainties denoting formal $1\sigma$ fit uncertainties obtained from the fit covariance.

To analyze the RB fidelity, we compare the primitive $\pi/2$-gate infidelity $\mathcal{I}=1-\mathcal{F}$ with the phase-noise limit $\mathcal{I}_{\mathrm{ph}}(f_{\mathrm{R}})$ obtained from simulations based on the measured qubit-frequency-noise PSDs using the experimental gate durations and sequence timings, following Ref.~\cite{Wu2025-highfidelity}. The excess infidelity is described as
\[
\mathcal{I}(f_{\mathrm{R}})
=\mathcal{I}_{\mathrm{ph}}(f_{\mathrm{R}})
+a_{\mathrm{d}}f_{\mathrm{R}}^{n_{\mathrm{RB}}}.
\]
We jointly fit $a_{\mathrm{d}}$ and $n_{\mathrm{RB}}$ to the Intel RB infidelities at $f_{\mathrm{R}}\geq1~\mathrm{MHz}$, minimizing the unweighted sum of squared residuals in $\log_{10}\mathcal{I}$ with $\mathcal{I}_{\mathrm{ph}}$ held fixed. The fit yields $n_{\mathrm{RB}}=2.46\pm0.28$, where the uncertainty is the formal $1\sigma$ standard error from the fit covariance.
We fix $n_{\mathrm{RB}}=2.46$ and fit $a_{\mathrm{d}}$ separately for the RIKEN and Intel data. The shaded bands in Fig.~\ref{fig:fig4}d propagate the uncertainty in $n_{\mathrm{RB}}$ by refitting $a_{\mathrm{d}}$ for each realization of the exponent and indicate the central 68\% interval of the resulting curves.

For small gate errors, the RB infidelity scales as $\Gamma_{\mathrm{RB}}t_{\mathrm{gate}}$. With $t_{\mathrm{gate}}\propto f_{\mathrm{R}}^{-1}$ and $P_{\mathrm{MW}}\propto f_{\mathrm{R}}^2$, the drive-dependent term corresponds to a rate-power exponent $\alpha_{\mathrm{RB}}=(n_{\mathrm{RB}}+1)/2=1.73\pm0.14$, whose central value is used in the RB-rate fit.

The transverse driving gradients in Fig.~\ref{fig:fig4}a were obtained from simulations of the respective micromagnet geometries. The RIKEN calculations follow Ref.~\cite{Wu2025-highfidelity}. For the Intel devices, we use a transverse-gradient range of $0.4\text{-}0.5~\mathrm{mT/nm}$ based on simulations provided by Intel, consistent with the range reported in Ref.~\cite{neyens_probing_2024}.

\paragraph{Microwave-power calibration.}
Throughout the manuscript, microwave powers denote nominal source-calibrated powers, rather than in-situ powers delivered to the qubit. Unless stated otherwise, microwave powers are reported in dBm and converted to milliwatts as $P_{\mathrm{mW}} = 10^{P_{\mathrm{dBm}}/10}$. The microwave transmission from the generator to the device depends on room-temperature electronics and wiring, cryostat cabling, attenuators, PCB transmission, and transmission on the device itself. A full in-situ calibration of the power delivered to each qubit is outside the scope of this work. Instead, we use a room-temperature spectrum analyzer calibration to define a consistent nominal power reference. The academic and industrial devices were measured in different cryogenic systems, but with the same nominal attenuation, the same PCB design, and comparable microwave cabling and cable lengths.

To obtain the power reference, a microwave switch placed after the microwave amplifier directed the same signal either to the device line or to a spectrum analyzer. In the spectrum analyzer path, we inserted an attenuation matching the nominal cryostat attenuation of the corresponding device line. This procedure defines the nominal powers quoted throughout the manuscript, but does not include additional losses between the room-temperature wiring and the qubit, such as frequency-dependent cable transmission, connector losses, PCB attenuation, device-package effects, or on-chip microwave transmission.

To assess whether such additional losses could explain the power differences in Fig.~\ref{fig:fig1}h, we performed room-temperature S-parameter measurements of the relevant microwave paths. These measurements are used only as a consistency check and not as a quantitative correction, because the transmission of the superconducting wiring improves at cryogenic temperatures, and the room-temperature measurements therefore do not directly represent the low-temperature transmission to the device. At the representative operating frequencies, the path used for the RIKEN measurements near $16.5~\mathrm{GHz}$ is approximately $3.3~\mathrm{dB}$ more attenuating at room temperature than the path used for the Intel measurements near $11.5~\mathrm{GHz}$. This additional wiring attenuation is not included in the nominal spectrum analyzer calibration. The measured transmission difference has the opposite sign to that required to explain the larger nominal powers of the Intel devices as a wiring artifact: at room temperature, the Intel path is less attenuating, yet the Intel qubits require higher nominal power to reach $f_{\mathrm{R}}=1~\mathrm{MHz}$ (Fig.~\ref{fig:fig1}h). Furthermore, the room-temperature S-parameter data exhibit a smooth frequency dependence, inconsistent with the large qubit-to-qubit spread in nominal excitation power observed in the academic-device data.

For the main cross-platform comparison, the representative simulated transverse magnetic-field gradients are $\approx 1.15~\mathrm{mT/nm}$ for the RIKEN device and $0.4\text{-}0.5~\mathrm{mT/nm}$ for the Intel device \cite{neyens_probing_2024}. For comparable orbital response to the microwave drive, this gradient difference implies that the Intel device would require a larger drive amplitude by a factor of $r=\nabla B_{\perp}^{\mathrm{RIKEN}}/\nabla B_{\perp}^{\mathrm{Intel}}\approx2.3\text{-}2.9$. Since microwave power scales with the square of drive amplitude, the predicted power increase is $10\log_{10}(r^2)=20\log_{10}r\approx7.2\text{-}9.2~\mathrm{dB}$. The measured median nominal-power difference of $7.7~\mathrm{dB}$ at $f_{\mathrm{R}}=1~\mathrm{MHz}$ between Intel Device 1 and the RIKEN reference falls within this predicted range, supporting the interpretation that reduced EDSR efficiency from weaker magnetic-field gradients is a major contributor to the larger nominal drive powers in the industrial devices.

\paragraph{Microwave-induced frequency shifts.}
Microwave-induced frequency shifts were measured using an off-resonant preheat protocol~\cite{Wu2025-highfidelity,Takeda2026_RB99999}. A microwave pulse of duration $t_{\mathrm{preheat}}$ was first applied at fixed power, after which the resonance frequency of an idling qubit was measured and compared with its value without preheating to extract the shift $\Delta f$. Repeating this procedure for different $t_{\mathrm{preheat}}$ yields the transient detuning shifts shown in Extended Data Fig.~\ref{fig:Fig_Ext_MWHeating}a. Measurements on both platforms were performed at an effective electron temperature of approximately 100~mK, similar to the conditions used for the RB experiments. This is relevant because the magnitude of the shift is temperature dependent and is known to decrease at elevated temperature as the available cooling power increases~\cite{undseth_hotter_2023,Takeda2026_RB99999}.

For the analysis in Extended Data Fig.~\ref{fig:Fig_Ext_MWHeating}, the measured shifts were summarized by their value at $t_{\mathrm{preheat}}=100~\mu\mathrm{s}$. Using the effective linear model $\Delta f(t)=r_{\mathrm{heat}}t$, these values define the measured detuning rates $r_{\mathrm{heat}}$, while RB simulations were used to determine the rates required to reproduce the primitive gate fidelities in Fig.~\ref{fig:fig4}d. Because the measured preheat traces are not strictly linear and often show partial saturation, this model is used only to define an effective detuning-rate metric over the operation-time window relevant for comparison with the RB simulations, not as a microscopic description of the full microwave-heating dynamics. The comparison in Extended Data Fig.~\ref{fig:Fig_Ext_MWHeating}c shows that the rates required to reproduce the RB loss are substantially larger than those measured directly.

\clearpage
\section*{Extended Data}\label{extended_data}
\renewcommand{\figurename}{Extended Data Figure}
\renewcommand{\tablename}{Extended Data Table}
\setcounter{figure}{0} 
\renewcommand{\theHfigure}{ED.\arabic{figure}}

\begin{figure*}[htbp]
\centering
\includegraphics[width=0.95\linewidth]{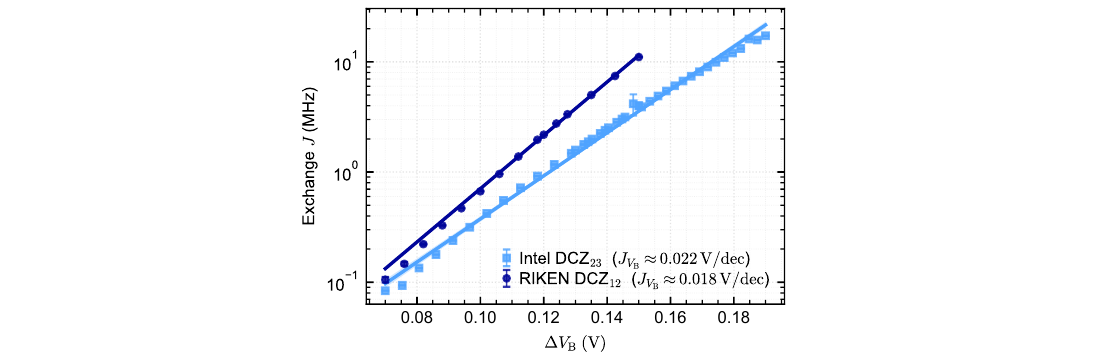}
\caption{\textbf{Inter-dot tunnel-barrier tunability.}
Exchange interaction $J$ as a function of barrier-gate voltage for representative qubit pairs in the RIKEN and Intel devices, extracted using a DCZ sequence. Solid lines are exponential fits, yielding characteristic voltage scales of $V_{\mathrm{dec}} = 0.0179~\mathrm{V}$ (RIKEN) and $0.0222~\mathrm{V}$ (Intel), corresponding to the voltage required to change $J$ by one decade. These voltage scales provide a compact measure of tunnel-barrier tunability and are set by the full electrostatic geometry, including gate layout, gate-quantum-well separation, and quantum-dot pitch. The exchange coupling tunability of the Intel device is found to be in reasonable agreement with 
Ref.~\cite{George2025-12qarrays}.
}
\label{fig:Fig_Ext_1_exchange}
\end{figure*}


\begin{figure*}[htbp]
\centering
\includegraphics[width=0.95\linewidth]{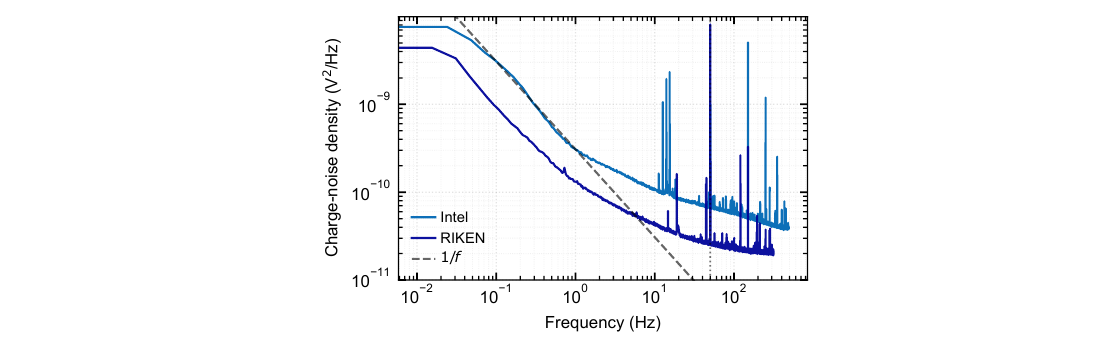}
\caption{Charge-noise power spectral densities measured in the industrial (Intel) and academic (RIKEN) devices, expressed in units of equivalent sensor plunger-gate voltage. The spectra are extracted from fluctuations of the charge-sensor signal during long repeated Ramsey measurements and converted to voltage units using the slope of the sensor quantum-dot Coulomb oscillations. For each platform, this conversion uses the corresponding single-shot sensor calibration and local sensor-slope measurement, so the spectra are plotted as equivalent plunger-gate voltage noise for the sensor used in that device. Presenting the data in plunger-gate voltage units provides a local measure of charge fluctuations in the gate-oxide environment sampled by the sensor. The industrial device exhibits a slightly higher sensor-referred charge-noise level across the measured frequency range, indicating that the reduced qubit-frequency noise observed in the main text arises from weaker qubit-frequency susceptibility rather than a quieter local charge-noise environment.
While the overall frequency dependence is similar in both devices, most noise peaks are specific to their respective experimental setups. A prominent peak at 50~Hz, arising from power-line interference, is visible in both devices. The gray dashed curve indicates a $1/f$ guide to the eye.
}
\label{fig:Fig_Ext_2_sensornoise}
\end{figure*}


\begin{figure*}[htbp]
\centering
\includegraphics[width=0.95\linewidth]{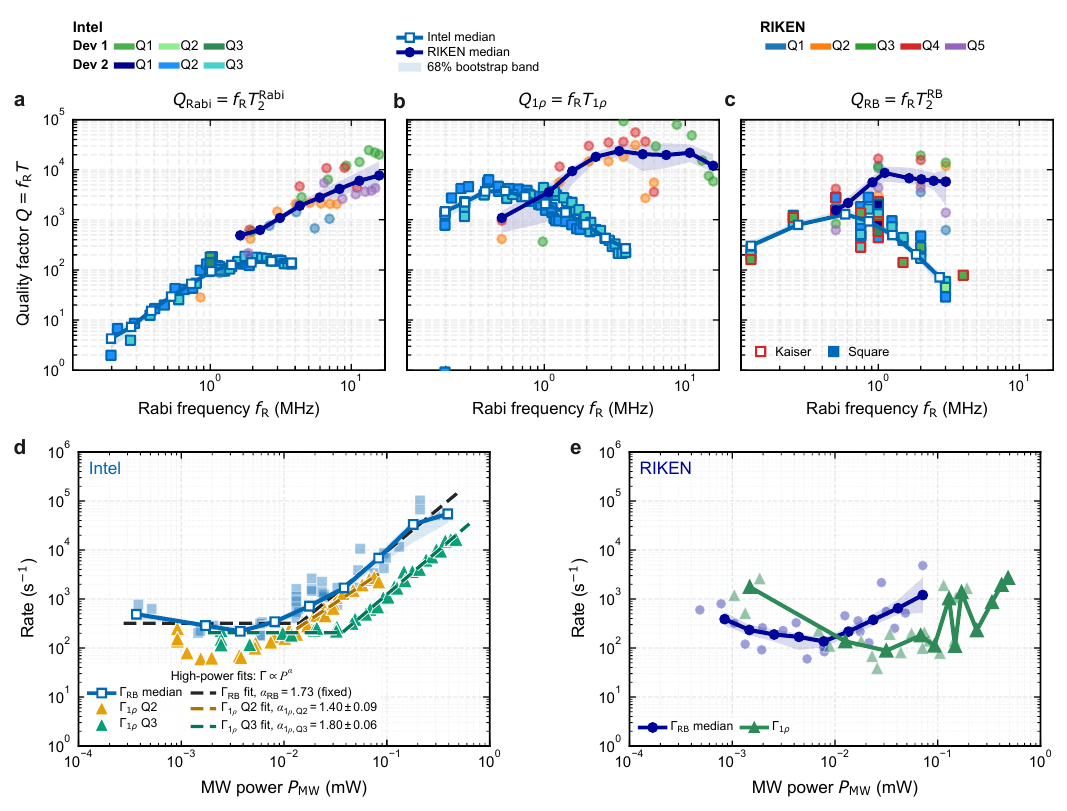}
\caption{\textbf{Driven-control quality factors and microwave-power-dependent decay rates.}
(a--c) Control-related quality factors as a function of Rabi frequency $f_{\mathrm{R}}$.
(a) Rabi quality factor $Q_{\mathrm{Rabi}}=f_{\mathrm{R}}T_{2}^{\mathrm{Rabi}}$, where $T_{2}^{\mathrm{Rabi}}$ is the decay time of resonantly driven Rabi oscillations.
(b) Spin-lock quality factor $Q_{1\rho}=f_{\mathrm{R}}T_{1\rho}$, where $T_{1\rho}$ is the rotating-frame relaxation time.
(c) RB quality factor $Q_{\mathrm{RB}}=f_{\mathrm{R}}T_{2}^{\mathrm{RB}}$, where $T_{2}^{\mathrm{RB}}=-t_{\mathrm{gate}}/\ln\mathcal{F}$ is the effective decay time obtained from the primitive $\pi/2$-gate fidelity $\mathcal{F}$.
Light points show individual datasets; dark filled circles and open squares with lines show the median trends for the RIKEN and Intel platforms, respectively. The RIKEN qubits attain higher peak quality factors in all three measures, while the Intel data suggest a smaller qubit-to-qubit spread in $Q_{\mathrm{Rabi}}$ and $Q_{\mathrm{RB}}$ over the measured Rabi-frequency ranges; no comparable reduction is resolved for $Q_{1\rho}$.
For the Intel RB data in (c), measurements performed using Kaiser-envelope pulses are indicated by red outlines; the remaining Intel measurements use square pulses. No systematic difference between the two pulse shapes is resolved over the measured operating range.
(d,e) RB-derived decay rates $\Gamma_{\mathrm{RB}}=1/T_{2}^{\mathrm{RB}}$ and spin-lock relaxation rates $\Gamma_{1\rho}=1/T_{1\rho}$ for (d) the Intel and (e) the RIKEN platform. In (d), the Intel $\Gamma_{\mathrm{RB}}$ data are combined into a shared trace across the measured qubits, whereas the $\Gamma_{1\rho}$ data are shown separately for Q2 and Q3. The rates are fitted using the model described in Methods. For the Intel RB rates, $\alpha_{\mathrm{RB}}=1.73$ is fixed from the infidelity fit in Fig.~\ref{fig:fig4}d, while the plateau rate and transition power are fitted. Independent spin-lock fits yield $\alpha_{1\rho,\mathrm{Q2}}=1.40\pm0.09$ and $\alpha_{1\rho,\mathrm{Q3}}=1.80\pm0.06$. Microwave powers are nominal source-calibrated values.
The Intel $Q_{\mathrm{Rabi}}$ data in (a) and the RB-derived data in (c,d) combine measurements from Devices~1 and~2, whereas the spin-lock data in (b,d) are from Device~2.
}
\label{fig:Fig_Ext_DrivenCoherences}
\end{figure*}


\begin{figure*}[htbp]
\centering
\includegraphics[width=0.95\linewidth]{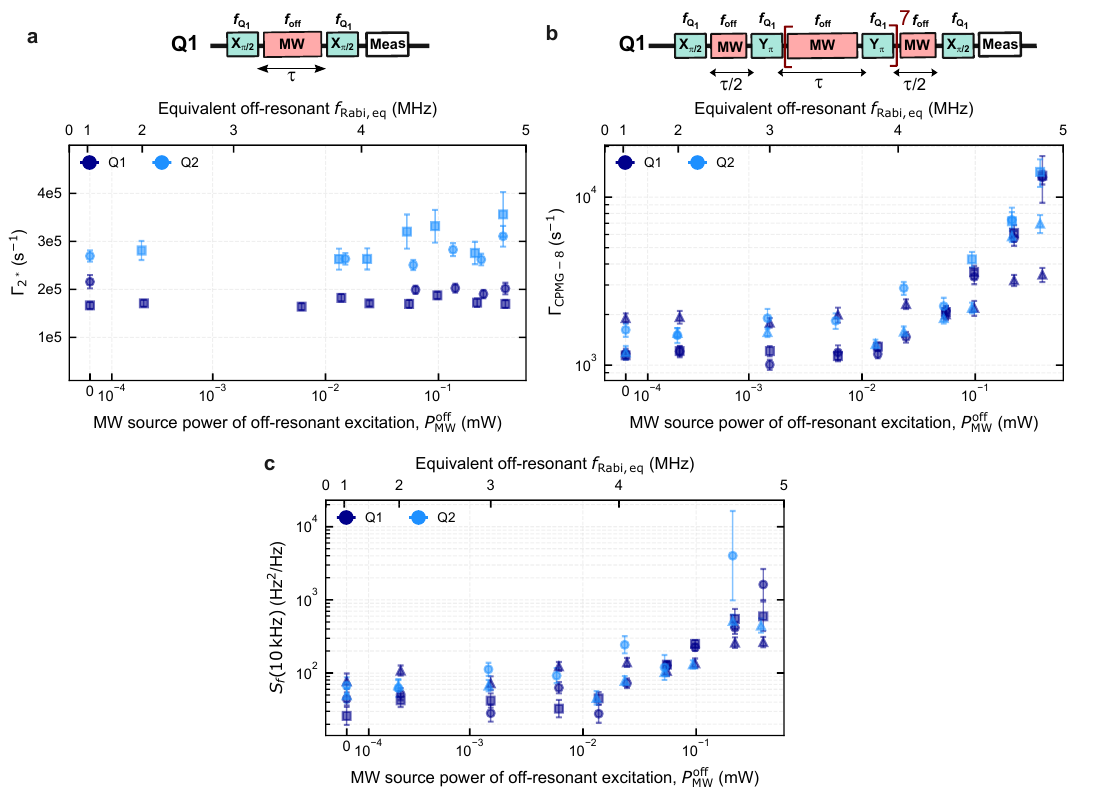}
\caption{\textbf{Microwave-induced dephasing.}
The measurements presented in this figure were performed on Intel Device~2.
(a) Ramsey dephasing rate $\Gamma_{2}^{*}=1/T_{2}^{*}$ measured with an off-resonant microwave tone applied during the free-evolution interval $\tau$, as illustrated in the schematic. The off-resonant tone is detuned by 40~MHz from the qubit resonance; the upper axes show the equivalent resonant Rabi frequency $f_{\mathrm{Rabi,eq}}$ obtained from the microwave-amplitude calibration. 
Ramsey is primarily sensitive to quasi-static and low-frequency qubit-frequency fluctuations and shows no clear systematic dependence on microwave power. The off-resonant tone probes microwave-induced dephasing without resonantly driving the qubit.
(b) CPMG-8 decay rate $\Gamma_{\mathrm{CPMG-8}}=1/T_{2}^{\mathrm{CPMG}}$ measured with off-resonant microwave irradiation applied during the free-evolution intervals. CPMG suppresses quasi-static fluctuations and shifts the measurement sensitivity towards the kHz frequency range set by its filter function; $\Gamma_{\mathrm{CPMG-8}}$ increases strongly with microwave power.
(c) Qubit-frequency-noise PSD $S_f(10\,\mathrm{kHz})$ extracted from the CPMG measurements, showing a corresponding increase with microwave power. Together with the power-dependent reduction of $T_{1\rho}$, which probes longitudinal noise near $f_{\mathrm{R}}$, these measurements are consistent with a microwave-induced noise contribution that becomes increasingly important at higher frequencies. However, the total microwave-irradiation time is substantially shorter in the Ramsey experiment than in the CPMG and spin-lock measurements. An irradiation-time-dependent response of the substrate, dielectric or gate stack, for example through heating or activation of fluctuators, could therefore also contribute to the weaker Ramsey response. The measurements do not distinguish between these two effects.
}
\label{fig:Fig_Ext_CPMG_Ramsey}
\end{figure*}

\begin{figure*}[htbp]
\centering
\includegraphics[width=0.95\linewidth]{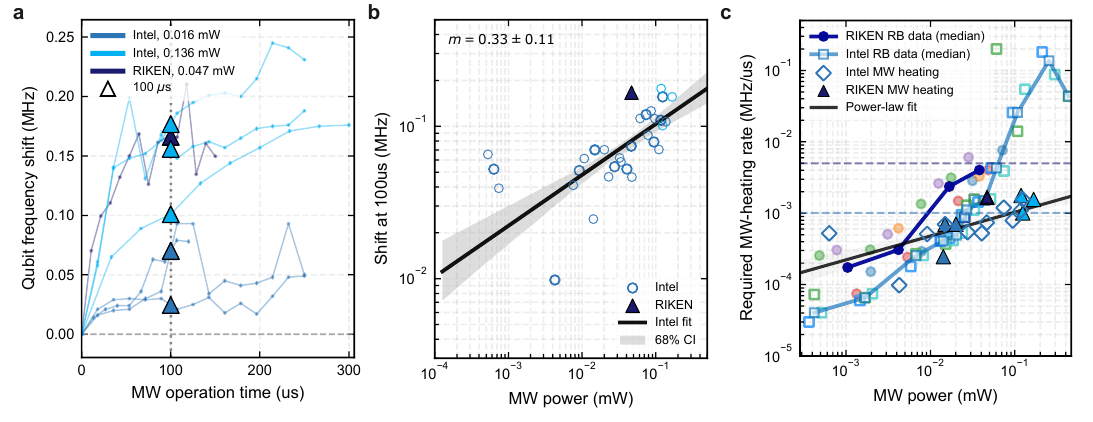}
\caption{\textbf{Measured MW-induced detuning shifts are insufficient to explain the observed drive-dependent loss in gate fidelity.}
(a) Representative qubit-frequency shifts versus off-resonant microwave preheat time for a RIKEN qubit and Intel Device~2. Triangles mark the shifts at 100~$\mu$s used in panel (b). Such shifts, often referred to as MW heating, have been reported previously in micromagnet-based spin-qubit devices~\cite{takeda_optimized_2018,undseth_hotter_2023,Wu2025-highfidelity,Takeda2026_RB99999}.
(b) Qubit-frequency shift at 100~$\mu$s as a function of microwave power, together with a power-law fit to the Intel data and its 68\% confidence interval. The RIKEN measurement lies within the same range.
(c) Effective MW-heating rates required to reproduce the measured primitive RB infidelities, compared with the rates inferred from the direct measurements in (a,b) and the corresponding power-law trend. For this comparison, the measured shifts are summarized by the effective linear model $\Delta f(t)=r_{\mathrm{heat}}t$, with $r_{\mathrm{heat}}$ defined by the shift at 100~$\mu$s. Because the measured transients often partially saturate, this linear model is used only as an effective detuning-rate metric rather than as a microscopic description of the heating dynamics. The rates required to reproduce the RB loss substantially exceed the directly measured rates at higher microwave powers, particularly for the industrial devices, showing that the measured MW-induced detuning shifts cannot account for the observed drive-dependent fidelity degradation within this effective model.
}
\label{fig:Fig_Ext_MWHeating}
\end{figure*}

\clearpage
\section*{Acknowledgements}
We thank the Intel team for providing the industrial Si/SiGe spin-qubit devices used in this work. We also thank R.~Kuroda for support with fabrication of the academic device, S.~Teraoka for technical support for the experimental setups, and J.~Rojas-Arias for assistance with the framework used to extract qubit-frequency noise power spectral densities and noise correlations.
This work was supported by JST Moonshot R\&D Grant Numbers JPMJMS226B and JPMJMS256H. L.C.C. acknowledges support from JSPS KAKENHI Grant Number 26K00642, RIKEN Incentive Research Project (202601080071). Y.H.W. acknowledges support from RIKEN’s IPA program. We also acknowledge support from JSPS KAKENHI Grant Numbers 22H01160 (T.K.) and 23H05455 (K.T.), 23K26483 (A.N.), JST PRESTO Grant Numbers JPMJPR2017 (T.N.) and JPMJPR23F8 (A.N.).

\section*{Author contributions}
L.C.C. and I.K.J. performed the experiments on the industrial devices. Y.H.W. and L.C.C. performed the experiments on the academic device. L.C.C. supervised the experiments. A.N. fabricated the academic device. K.T., A.N., T.K., and T.N. contributed to the data acquisition and discussed the results. G.S. grew, developed, and supplied the silicon-28/silicon-germanium heterostructure for the academic device. L.C.C. wrote the manuscript with input from all co-authors. S.T. supervised the project.

\section*{Competing interests}
The authors declare no competing interests.

\section*{Data availability}
The data that support the findings of this study are openly available in a Zenodo repository \cite{camenzind_intel_zenodo}.

\section*{Use of artificial intelligence tools}
AI tools (ChatGPT and Codex) assisted with manuscript preparation and development of analysis and plotting code. The authors reviewed and verified all outputs.

\clearpage
\bibliographystyle{sn-standardnature-10authors}
\bibliography{sn-bibliography}

@article{yoneda_robust_2015,
  title = {Robust micromagnet design for fast electrical manipulations of single spins in quantum dots},
  volume = {8},
  url = {http://dx.doi.org/10.7567/APEX.8.084401},
  DOI = {10.7567/apex.8.084401},
  number = {8},
  journal = {Applied Physics Express},
  publisher = {IOP Publishing},
  author = {Yoneda,  Jun and Otsuka,  Tomohiro and Takakura,  Tatsuki and Pioro-Ladrière,  Michel and Brunner,  Roland and Lu,  Hong and Nakajima,  Takashi and Obata,  Toshiaki and Noiri,  Akito and Palmstrøm,  Christopher J. and Gossard,  Arthur C. and Tarucha,  Seigo},
  year = {2015},
  month = jul,
  pages = {084401}
}

@article{nakajima_coherence_2020,
  title = {Coherence of a Driven Electron Spin Qubit Actively Decoupled from Quasistatic Noise},
  volume = {10},
  url = {http://dx.doi.org/10.1103/PhysRevX.10.011060},
  DOI = {10.1103/physrevx.10.011060},
  number = {1},
  journal = {Physical Review X},
  publisher = {American Physical Society (APS)},
  author = {Nakajima,  Takashi and Noiri,  Akito and Kawasaki,  Kento and Yoneda,  Jun and Stano,  Peter and Amaha,  Shinichi and Otsuka,  Tomohiro and Takeda,  Kenta and Delbecq,  Matthieu R. and Allison,  Giles and Ludwig,  Arne and Wieck,  Andreas D. and Loss,  Daniel and Tarucha,  Seigo},
  year = {2020},
  month = mar
}

@article{boter_spiderweb_2022,
  title = {Spiderweb array: {A} sparse spin-qubit array},
  volume = {18},
  doi = {10.1103/PhysRevApplied.18.024053},
  number = {2},
  journal = {Phys. Rev. Appl.},
  author = {Boter, Jelmer M. and Dehollain, Juan P. and van Dijk, Jeroen P. G. and Xu, Yuanxing and Hensgens, Toivo and Versluis, Richard and Naus, Henricus W. L. and Clarke, James S. and Veldhorst, Menno and Sebastiano, Fabio and Vandersypen, Lieven M. K.},
  month = aug,
  year = {2022},
  note = {arXiv:2110.00189 [cond-mat, physics:quant-ph]},
  pages = {024053}
}

@article{camenzind_spin_2022,
  title = {A spin qubit in a fin field-effect transistor},
  volume = {5},
  url = {https://www.nature.com/articles/s41928-022-00722-0},
  doi = {10.1038/s41928-022-00722-0},
  number = {3},
  journal = {Nat. Electron.},
  author = {Camenzind, Leon C. and Geyer, Simon and Fuhrer, Andreas and Warburton, Richard J. and Zumbühl, Dominik M. and Kuhlmann, Andreas V.},
  month = mar,
  year = {2022},
  pages = {178--183}
}

@article{yang_operation_2020,
  title = {Operation of a silicon quantum processor unit cell above one kelvin},
  volume = {580},
  url = {http://www.nature.com/articles/s41586-020-2171-6},
  doi = {10.1038/s41586-020-2171-6},
  number = {7803},
  journal = {Nature},
  author = {Yang, C. H. and Leon, R. C. C. and Hwang, J. C. C. and Saraiva, A. and Tanttu, T. and Huang, W. and Camirand Lemyre, J. and Chan, K. W. and Tan, K. Y. and Hudson, F. E. and Itoh, K. M. and Morello, A. and Pioro-Ladrière, M. and Laucht, A. and Dzurak, A. S.},
  month = apr,
  year = {2020},
  pages = {350--354}
}

@article{fowler_surface_2012,
  title = {Surface codes: {Towards} practical large-scale quantum computation},
  volume = {86},
  url = {https://link.aps.org/doi/10.1103/PhysRevA.86.032324},
  doi = {10.1103/PhysRevA.86.032324},
  number = {3},
  journal = {Phys. Rev. A},
  author = {Fowler, Austin G. and Mariantoni, Matteo and Martinis, John M. and Cleland, Andrew N.},
  month = sep,
  year = {2012},
  pages = {032324}
}

@article{loss_quantum_1998,
  title = {Quantum computation with quantum dots},
  volume = {57},
  url = {https://link.aps.org/doi/10.1103/PhysRevA.57.120},
  doi = {10.1103/PhysRevA.57.120},
  number = {1},
  journal = {Phys. Rev. A},
  author = {Loss, Daniel and DiVincenzo, David P.},
  month = jan,
  year = {1998},
  pages = {120--126}
}

@article{noiri_fast_2022,
  title = {Fast universal quantum gate above the fault-tolerance threshold in silicon},
  volume = {601},
  url = {https://www.nature.com/articles/s41586-021-04182-y},
  doi = {10.1038/s41586-021-04182-y},
  number = {7893},
  journal = {Nature},
  author = {Noiri, Akito and Takeda, Kenta and Nakajima, Takashi and Kobayashi, Takashi and Sammak, Amir and Scappucci, Giordano and Tarucha, Seigo},
  month = jan,
  year = {2022},
  pages = {338--342}
}

@article{petit_universal_2020,
  title = {Universal quantum logic in hot silicon qubits},
  volume = {580},
  url = {http://www.nature.com/articles/s41586-020-2170-7},
  doi = {10.1038/s41586-020-2170-7},
  number = {7803},
  journal = {Nature},
  author = {Petit, L. and Eenink, H. G. J. and Russ, M. and Lawrie, W. I. L. and Hendrickx, N. W. and Philips, S. G. J. and Clarke, J. S. and Vandersypen, L. M. K. and Veldhorst, M.},
  month = apr,
  year = {2020},
  pages = {355--359}
}

@article{takeda_fault-tolerant_2016,
  title = {A fault-tolerant addressable spin qubit in a natural silicon quantum dot},
  volume = {2},
  url = {https://www.science.org/doi/10.1126/sciadv.1600694},
  doi = {10.1126/sciadv.1600694},
  number = {8},
  journal = {Sci. Adv.},
  author = {Takeda, Kenta and Kamioka, Jun and Otsuka, Tomohiro and Yoneda, Jun and Nakajima, Takashi and Delbecq, Matthieu R. and Amaha, Shinichi and Allison, Giles and Kodera, Tetsuo and Oda, Shunri and Tarucha, Seigo},
  month = aug,
  year = {2016},
  pages = {e1600694}
}

@article{takeda_optimized_2018,
  title = {Optimized electrical control of a {Si}/{SiGe} spin qubit in the presence of an induced frequency shift},
  volume = {4},
  url = {http://www.nature.com/articles/s41534-018-0105-z},
  doi = {10.1038/s41534-018-0105-z},
  number = {1},
  journal = {Npj Quantum Inf.},
  author = {Takeda, K. and Yoneda, J. and Otsuka, T. and Nakajima, T. and Delbecq, M. R. and Allison, G. and Hoshi, Y. and Usami, N. and Itoh, K. M. and Oda, S. and Kodera, T. and Tarucha, S.},
  month = dec,
  year = {2018},
  pages = {54}
}

@article{veldhorst_silicon_2017,
  title = {Silicon {CMOS} architecture for a spin-based quantum computer},
  volume = {8},
  url = {http://www.nature.com/articles/s41467-017-01905-6},
  doi = {10.1038/s41467-017-01905-6},
  number = {1},
  journal = {Nat. Commun.},
  author = {Veldhorst, M. and Eenink, H. G. J. and Yang, C. H. and Dzurak, A. S.},
  month = dec,
  year = {2017},
  pages = {1766}
}

@article{ha_flexible_2022,
  title = {A flexible design platform for {Si}/{SiGe} exchange-only qubits with low disorder},
  volume = {22},
  doi = {10.1021/acs.nanolett.1c03026},
  number = {3},
  journal = {Nano Lett.},
  author = {Ha, Wonill and Ha, Sieu D. and Choi, Maxwell D. and Tang, Yan and Schmitz, Adele E. and Levendorf, Mark P. and Lee, Kangmu and Chappell, James M. and Adams, Tower S. and Hulbert, Daniel R. and Acuna, Edwin and Noah, Ramsey S. and Matten, Justine W. and Jura, Michael P. and Wright, Jeffrey A. and Rakher, Matthew T. and Borselli, Matthew G.},
  month = feb,
  year = {2022},
  pages = {1443--1448}
}

@article{xue_cmos-based_2021,
  title = {{CMOS}-based cryogenic control of silicon quantum circuits},
  volume = {593},
  url = {http://www.nature.com/articles/s41586-021-03469-4},
  doi = {10.1038/s41586-021-03469-4},
  number = {7858},
  journal = {Nature},
  author = {Xue, Xiao and Patra, Bishnu and van Dijk, Jeroen P. G. and Samkharadze, Nodar and Subramanian, Sushil and Corna, Andrea and Paquelet Wuetz, Brian and Jeon, Charles and Sheikh, Farhana and Juarez-Hernandez, Esdras and Esparza, Brando Perez and Rampurawala, Huzaifa and Carlton, Brent and Ravikumar, Surej and Nieva, Carlos and Kim, Sungwon and Lee, Hyung-Jin and Sammak, Amir and Scappucci, Giordano and Veldhorst, Menno and Sebastiano, Fabio and Babaie, Masoud and Pellerano, Stefano and Charbon, Edoardo and Vandersypen, Lieven M. K.},
  month = may,
  year = {2021},
  pages = {205--210}
}

@article{xue_quantum_2022,
  title = {Quantum logic with spin qubits crossing the surface code threshold},
  volume = {601},
  url = {https://www.nature.com/articles/s41586-021-04273-w},
  doi = {10.1038/s41586-021-04273-w},
  number = {7893},
  journal = {Nature},
  author = {Xue, Xiao and Russ, Maximilian and Samkharadze, Nodar and Undseth, Brennan and Sammak, Amir and Scappucci, Giordano and Vandersypen, Lieven M. K.},
  month = jan,
  year = {2022},
  pages = {343--347}
}

@article{yoneda_quantum-dot_2018,
  title = {A quantum-dot spin qubit with coherence limited by charge noise and fidelity higher than 99.9\%},
  volume = {13},
  url = {http://www.nature.com/articles/s41565-017-0014-x},
  doi = {10.1038/s41565-017-0014-x},
  number = {2},
  journal = {Nat. Nanotechnol.},
  author = {Yoneda, Jun and Takeda, Kenta and Otsuka, Tomohiro and Nakajima, Takashi and Delbecq, Matthieu R. and Allison, Giles and Honda, Takumu and Kodera, Tetsuo and Oda, Shunri and Hoshi, Yusuke and Usami, Noritaka and Itoh, Kohei M. and Tarucha, Seigo},
  month = feb,
  year = {2018},
  pages = {102--106}
}

@article{neyens_probing_2024,
  title = {Probing single electrons across 300-mm spin qubit wafers},
  volume = {629},
  url = {http://dx.doi.org/10.1038/s41586-024-07275-6},
  DOI = {10.1038/s41586-024-07275-6},
  number = {8010},
  journal = {Nature},
  publisher = {Springer Science and Business Media LLC},
  author = {Neyens,  Samuel and Zietz,  Otto K. and Watson,  Thomas F. and Luthi,  Florian and Nethwewala,  Aditi and George,  Hubert C. and Henry,  Eric and Islam,  Mohammad and Wagner,  Andrew J. and Borjans,  Felix and Connors,  Elliot J. and Corrigan,  J. and Curry,  Matthew J. and Keith,  Daniel and Kotlyar,  Roza and Lampert,  Lester F. and Mądzik,  Mateusz T. and Millard,  Kent and Mohiyaddin,  Fahd A. and Pellerano,  Stefano and Pillarisetty,  Ravi and Ramsey,  Mick and Savytskyy,  Rostyslav and Schaal,  Simon and Zheng,  Guoji and Ziegler,  Joshua and Bishop,  Nathaniel C. and Bojarski,  Stephanie and Roberts,  Jeanette and Clarke,  James S.},
  year = {2024},
  month = may,
  pages = {80–85}
}

@article{undseth_hotter_2023,
  title = "Hotter is easier: Unexpected temperature dependence of spin
               qubit frequencies",
  author = {Undseth, Brennan and Pietx-Casas, Oriol and Raymenants, Eline and Mehmandoost, Mohammad and M{\k{a}}dzik, Mateusz T. and Philips, Stephan G. J. and {de Snoo}, Sander L. and Michalak, David J. and Amitonov, Sergey V. and Tryputen, Larysa and Wuetz, Brian Paquelet and Fezzi, Viviana and {Degli Esposti}, Davide and Sammak, Amir and Scappucci, Giordano and Vandersypen, Lieven M. K.},
  journal = "Phys. Rev. X.",
  publisher = "American Physical Society (APS)",
  volume = 13,
  number = 4,
  month = oct,
  year = 2023,
  pages = {041015},
  doi = {10.1103/PhysRevX.13.041015}
}

@article{Bartee2025,
  title = {Spin-qubit control with a milli-kelvin CMOS chip},
  volume = {643},
  url = {http://dx.doi.org/10.1038/s41586-025-09157-x},
  DOI = {10.1038/s41586-025-09157-x},
  number = {8071},
  journal = {Nature},
  publisher = {Springer Science and Business Media LLC},
  author = {Bartee,  Samuel K. and Gilbert,  Will and Zuo,  Kun and Das,  Kushal and Tanttu,  Tuomo and Yang,  Chih Hwan and Dumoulin Stuyck,  Nard and Pauka,  Sebastian J. and Su,  Rocky Y. and Lim,  Wee Han and Serrano,  Santiago and Escott,  Christopher C. and Hudson,  Fay E. and Itoh,  Kohei M. and Laucht,  Arne and Dzurak,  Andrew S. and Reilly,  David J.},
  year = {2025},
  month = jun,
  pages = {382–387}
}

@article{philips_universal_2022,
  title = {Universal control of a six-qubit quantum processor in silicon},
  volume = {609},
  url = {https://www.nature.com/articles/s41586-022-05117-x},
  doi = {10.1038/s41586-022-05117-x},
  number = {7929},
  journal = {Nature},
  author = {Philips, Stephan G. J. and Mądzik, Mateusz T. and Amitonov, Sergey V. and {de Snoo}, Sander L. and Russ, Maximilian and Kalhor, Nima and Volk, Christian and Lawrie, William I. L. and Brousse, Delphine and Tryputen, Larysa and Wuetz, Brian Paquelet and Sammak, Amir and Veldhorst, Menno and Scappucci, Giordano and Vandersypen, Lieven M. K.},
  month = sep,
  year = {2022},
  pages = {919--924}
}

@article{mills_two-qubit_2022,
  title = {Two-qubit silicon quantum processor with operation fidelity exceeding 99\%},
  volume = {8},
  url = {https://www.science.org/doi/10.1126/sciadv.abn5130},
  doi = {10.1126/sciadv.abn5130},
  number = {14},
  journal = {Sci. Adv.},
  author = {Mills, Adam R. and Guinn, Charles R. and Gullans, Michael J. and Sigillito, Anthony J. and Feldman, Mayer M. and Nielsen, Erik and Petta, Jason R.},
  month = apr,
  year = {2022},
  pages = {eabn5130}
}

@article{takeda_quantum_2022,
  title = {Quantum error correction with silicon spin qubits},
  volume = {608},
  url = {https://www.nature.com/articles/s41586-022-04986-6},
  doi = {10.1038/s41586-022-04986-6},
  number = {7924},
  journal = {Nature},
  author = {Takeda, Kenta and Noiri, Akito and Nakajima, Takashi and Kobayashi, Takashi and Tarucha, Seigo},
  month = aug,
  year = {2022},
  pages = {682--686}
}

@article{zwerver_qubits_2022,
  title = {Qubits made by advanced semiconductor manufacturing},
  volume = {5},
  url = {https://www.nature.com/articles/s41928-022-00727-9},
  doi = {10.1038/s41928-022-00727-9},
  number = {3},
  journal = {Nat. Electron.},
  author = {Zwerver, A. M. J. and Krähenmann, T. and Watson, T. F. and Lampert, L. and George, H. C. and Pillarisetty, R. and Bojarski, S. A. and Amin, P. and Amitonov, S. V. and Boter, J. M. and Caudillo, R. and Correas-Serrano, D. and Dehollain, J. P. and Droulers, G. and Henry, E. M. and Kotlyar, R. and Lodari, M. and Lüthi, F. and Michalak, D. J. and Mueller, B. K. and Neyens, S. and Roberts, J. and Samkharadze, N. and Zheng, G. and Zietz, O. K. and Scappucci, G. and Veldhorst, M. and Vandersypen, L. M. K. and Clarke, J. S.},
  month = mar,
  year = {2022},
  pages = {184--190}
}

@article{yoneda_noise-correlation_2023,
  title = {Noise-correlation spectrum for a pair of spin qubits in silicon},
  volume = {19},
  pages = {1793–1798},
  url = {https://www.nature.com/articles/s41567-023-02238-6},
  doi = {10.1038/s41567-023-02238-6},
  journal = {Nat. Phys.},
  author = {Yoneda, J. and Rojas-Arias, J. S. and Stano, P. and Takeda, K. and Noiri, A. and Nakajima, T. and Loss, D. and Tarucha, S.},
  month = oct,
  year = {2023}
}

@article{kawakami_electrical_2014,
  title = {Electrical control of a long-lived spin qubit in a {Si}/{SiGe} quantum dot},
  volume = {9},
  url = {http://www.nature.com/articles/nnano.2014.153},
  doi = {10.1038/nnano.2014.153},
  number = {9},
  journal = {Nat. Nanotechnol.},
  author = {Kawakami, E. and Scarlino, P. and Ward, D. R. and Braakman, F. R. and Savage, D. E. and Lagally, M. G. and Friesen, Mark and Coppersmith, S. N. and Eriksson, M. A. and Vandersypen, L. M. K.},
  month = sep,
  year = {2014},
  pages = {666--670}
}

@Article{Vandersypen_interfacing_2017,
  author = {L. M. K. Vandersypen and H. Bluhm and J. S. Clarke and A. S. Dzurak and R. Ishihara and A. Morello and D. J. Reilly and L. R. Schreiber and M. Veldhorst},
  journal = {Npj Quantum Inf.},
  title = {Interfacing spin qubits in quantum dots and donors{\textemdash}hot, dense, and coherent},
  year = {2017},
  month = sep,
  number = {1},
  volume = {3},
  pages = {34},
  doi = {10.1038/s41534-017-0038-y},
  publisher = {Springer Science and Business Media {LLC}},
  url = {https://doi.org/10.1038/s41534-017-0038-y}
}

@Article{Pioro-Ladriere_2008,
  title = {Electrically driven single-electron spin resonance in a slanting Zeeman field},
  volume = {4},
  DOI = {10.1038/nphys1053},
  number = {10},
  journal = {Nature Physics},
  author = {Pioro-Ladrière, M. and Obata, T. and Tokura, Y. and Shin, Y.-S. and Kubo, T. and Yoshida, K. and Taniyama, T. and Tarucha, S.},
  year = {2008},
  month = aug,
  pages = {776–779}
}

@ARTICLE{George2025-12qarrays,
  title = "12-spin-qubit arrays fabricated on a 300 mm semiconductor
               manufacturing line",
  author = "George, Hubert C and M{\k a}dzik, Mateusz T and Henry, Eric M
               and Wagner, Andrew J and Islam, Mohammad M and Borjans, Felix
               and Connors, Elliot J and Corrigan, J and Curry, Matthew and
               Harper, Michael K and Keith, Daniel and Lampert, Lester and
               Luthi, Florian and Mohiyaddin, Fahd A and Murcia, Sandra and
               Nair, Rohit and Nahm, Rambert and Nethwewala, Aditi and Neyens,
               Samuel and Patra, Bishnu and Raharjo, Roy D and Rogan, Carly and
               Savytskyy, Rostyslav and Watson, Thomas F and Ziegler, Josh and
               Zietz, Otto K and Pellerano, Stefano and Pillarisetty, Ravi and
               Bishop, Nathaniel C and Bojarski, Stephanie A and Roberts,
               Jeanette and Clarke, James S",
  journal = "Nano Lett.",
  publisher = "American Chemical Society (ACS)",
  volume = 25,
  number = 2,
  pages = "793--799",
  month = jan,
  year = 2025,
  doi = {10.1021/acs.nanolett.4c05205}
}

@ARTICLE{Gidney2021-lk,
  title = "How to factor 2048 bit {RSA} integers in 8 hours using 20
               million noisy qubits",
  author = "Gidney, Craig and Eker{\aa}, Martin",
  journal = "Quantum",
  publisher = "Verein zur Forderung des Open Access Publizierens in den
               Quantenwissenschaften",
  volume = 5,
  number = 433,
  pages = "433",
  month = apr,
  year = 2021
}

@ARTICLE{Dowling2003-sx,
  title = "Quantum technology: the second quantum revolution",
  author = "Dowling, Jonathan P and Milburn, Gerard J",
  journal = "Philos. Trans. A Math. Phys. Eng. Sci.",
  publisher = "The Royal Society",
  volume = 361,
  number = 1809,
  pages = "1655--1674",
  month = aug,
  year = 2003,
  doi = {10.1098/rsta.2003.1227}
}

@ARTICLE{Abrams1999-lc,
  title = "Quantum algorithm providing exponential speed increase for
               finding eigenvalues and eigenvectors",
  author = "Abrams, Daniel S and Lloyd, Seth",
  journal = "Phys. Rev. Lett.",
  publisher = "American Physical Society (APS)",
  volume = 83,
  number = 24,
  pages = "5162--5165",
  month = dec,
  year = 1999,
  doi = {10.1103/PhysRevLett.83.5162}
}

@ARTICLE{Berry2019-md,
  title = "Qubitization of arbitrary basis quantum chemistry leveraging
               sparsity and low rank factorization",
  author = "Berry, Dominic W and Gidney, Craig and Motta, Mario and McClean,
               Jarrod R and Babbush, Ryan",
  journal = "Quantum",
  publisher = "Verein zur Forderung des Open Access Publizierens in den
               Quantenwissenschaften",
  volume = 3,
  number = 208,
  pages = "208",
  month = dec,
  year = 2019,
  doi = {10.22331/q-2019-12-02-208}
}

@ARTICLE{Ladd2010-tf,
  title = "Quantum computers",
  author = "Ladd, T D and Jelezko, F and Laflamme, R and Nakamura, Y and
               Monroe, C and O'Brien, J L",
  journal = "Nature",
  publisher = "Springer Science and Business Media LLC",
  volume = 464,
  number = 7285,
  pages = "45--53",
  month = mar,
  year = 2010,
  doi = {10.1038/nature08812}
}

@ARTICLE{Feynman1982-wh,
  title = "Simulating physics with computers",
  author = "Feynman, Richard P",
  journal = "Int. J. Theor. Phys.",
  publisher = "Springer Science and Business Media LLC",
  volume = 21,
  number = "6-7",
  pages = "467--488",
  month = jun,
  year = 1982,
  doi = {10.1007/BF02650179}
}

@ARTICLE{Farhi2022-uv,
  title = "The Quantum Approximate Optimization Algorithm and the
               {Sherrington-Kirkpatrick} model at infinite size",
  author = "Farhi, Edward and Goldstone, Jeffrey and Gutmann, Sam and Zhou,
               Leo",
  journal = "Quantum",
  publisher = "Verein zur Forderung des Open Access Publizierens in den
               Quantenwissenschaften",
  volume = 6,
  number = 759,
  pages = "759",
  month = jul,
  year = 2022,
  doi = {10.22331/q-2022-07-07-759}
}

@ARTICLE{Li2025-yd,
  title = "A trilinear quantum dot architecture for semiconductor spin
                  qubits",
  author = "Li, R and Levajac, V and Godfrin, C and Kubicek, S and
                  Simion, G and Raes, B and Beyne, S and Fattal, I and
                  Loenders, A and De Roeck, W and Mongillo, M and Wan, D and De
                  Greve, K",
  year = 2025,
  primaryClass = "quant-ph",
  eprint = "2501.17814"
}

@ARTICLE{Takeda2024-fm,
  title = "Rapid single-shot parity spin readout in a silicon double
               quantum dot with fidelity exceeding 99\%",
  author = "Takeda, Kenta and Noiri, Akito and Nakajima, Takashi and
               Camenzind, Leon C and Kobayashi, Takashi and Sammak, Amir and
               Scappucci, Giordano and Tarucha, Seigo",
  journal = "Npj Quantum Inf.",
  publisher = "Springer Science and Business Media LLC",
  volume = 10,
  number = 1,
  month = feb,
  year = 2024,
  doi = {10.1038/s41534-024-00813-0}
}

@ARTICLE{Wu2025-highfidelity,
  title = "Simultaneous high-fidelity single-qubit gates in a spin
                   qubit array",
  author = "Wu, Yi-Hsien and Camenzind, Leon C and B{\"u}tler, Patrick
                   and Jin, Ik Kyeong and Noiri, Akito and Takeda, Kenta and
                   Nakajima, Takashi and Kobayashi, Takashi and Scappucci,
                   Giordano and Goan, Hsi-Sheng and Tarucha, Seigo",
  url = {http://dx.doi.org/10.1038/s41567-026-03466-2},
  DOI = {10.1038/s41567-026-03466-2},
  journal = {Nature Physics},
  publisher = {Springer Science and Business Media LLC},
  year = {2026},
  month = Oct
}

@article{Huang2024,
  title = {High-fidelity spin qubit operation and algorithmic initialization above 1 K},
  volume = {627},
  url = {http://dx.doi.org/10.1038/s41586-024-07160-2},
  DOI = {10.1038/s41586-024-07160-2},
  number = {8005},
  journal = {Nature},
  publisher = {Springer Science and Business Media LLC},
  author = {Huang,  Jonathan Y. and Su,  Rocky Y. and Lim,  Wee Han and Feng,  MengKe and van Straaten,  Barnaby and Severin,  Brandon and Gilbert,  Will and Dumoulin Stuyck,  Nard and Tanttu,  Tuomo and Serrano,  Santiago and Cifuentes,  Jesus D. and Hansen,  Ingvild and Seedhouse,  Amanda E. and Vahapoglu,  Ensar and Leon,  Ross C. C. and Abrosimov,  Nikolay V. and Pohl,  Hans-Joachim and Thewalt,  Michael L. W. and Hudson,  Fay E. and Escott,  Christopher C. and Ares,  Natalia and Bartlett,  Stephen D. and Morello,  Andrea and Saraiva,  Andre and Laucht,  Arne and Dzurak,  Andrew S. and Yang,  Chih Hwan},
  year = {2024},
  month = mar,
  pages = {772–777}
}

@article{Steinacker2025,
  title = {Industry-compatible silicon spin-qubit unit cells exceeding 99\% fidelity},
  volume = {646},
  url = {http://dx.doi.org/10.1038/s41586-025-09531-9},
  DOI = {10.1038/s41586-025-09531-9},
  number = {8083},
  journal = {Nature},
  publisher = {Springer Science and Business Media LLC},
  author = {Steinacker,  Paul and Dumoulin Stuyck,  Nard and Lim,  Wee Han and Tanttu,  Tuomo and Feng,  MengKe and Serrano,  Santiago and Nickl,  Andreas and Candido,  Marco and Cifuentes,  Jesus D. and Vahapoglu,  Ensar and Bartee,  Samuel K. and Hudson,  Fay E. and Chan,  Kok Wai and Kubicek,  Stefan and Jussot,  Julien and Canvel,  Yann and Beyne,  Sofie and Shimura,  Yosuke and Loo,  Roger and Godfrin,  Clement and Raes,  Bart and Baudot,  Sylvain and Wan,  Danny and Laucht,  Arne and Yang,  Chih Hwan and Saraiva,  Andre and Escott,  Christopher C. and De Greve,  Kristiaan and Dzurak,  Andrew S.},
  year = {2025},
  month = sep,
  pages = {81–87}
}

@article{deLeon2021,
  title = {Materials challenges and opportunities for quantum computing hardware},
  volume = {372},
  url = {http://dx.doi.org/10.1126/science.abb2823},
  DOI = {10.1126/science.abb2823},
  number = {6539},
  journal = {Science},
  publisher = {American Association for the Advancement of Science (AAAS)},
  author = {de Leon,  Nathalie P. and Itoh,  Kohei M. and Kim,  Dohun and Mehta,  Karan K. and Northup,  Tracy E. and Paik,  Hanhee and Palmer,  B. S. and Samarth,  N. and Sangtawesin,  Sorawis and Steuerman,  D. W.},
  year = {2021},
  month = apr
}

@ARTICLE{RojasArias2026Scaling,
  title = "Scaling of silicon spin qubits under correlated noise",
  author = "Rojas-Arias, Juan S and Camenzind, Leon C and Wu, Yi-Hsien
                   and Stano, Peter and Noiri, Akito and Takeda, Kenta and
                   Nakajima, Takashi and Kobayashi, Takashi and Scappucci,
                   Giordano and Loss, Daniel and Tarucha, Seigo",
  month = mar,
  year = 2026,
  archivePrefix = "arXiv",
  primaryClass = "cond-mat.mes-hall",
  eprint = "2603.03051"
}

@article{DiVincenzo2000,
  author = {DiVincenzo, David P.},
  title = {The Physical Implementation of Quantum Computation},
  journal = {Fortschritte der Physik},
  volume = {48},
  number = {9-11},
  pages = {771-783},
  doi = {10.1002/1521-3978(200009)48:9/11<771::AID-PROP771>3.0.CO;2-E},
  year = {2000}
}

@article{Wu2024,
  title = {Hamiltonian phase error in resonantly driven CNOT gate above the fault-tolerant threshold},
  volume = {10},
  url = {http://dx.doi.org/10.1038/s41534-023-00802-9},
  DOI = {10.1038/s41534-023-00802-9},
  number = {1},
  journal = {npj Quantum Information},
  publisher = {Springer Science and Business Media LLC},
  author = {Wu,  Yi-Hsien and Camenzind,  Leon C. and Noiri,  Akito and Takeda,  Kenta and Nakajima,  Takashi and Kobayashi,  Takashi and Chang,  Chien-Yuan and Sammak,  Amir and Scappucci,  Giordano and Goan,  Hsi-Sheng and Tarucha,  Seigo},
  year = {2024},
  month = jan
}

@ARTICLE{Takeda2026_RB99999,
  title = "Assessing fidelity-limiting factors and achieving
                   single-qubit gate fidelity beyond 99.999\% in driven silicon
                   spin qubits",
  author = "Takeda, Kenta and Noiri, Akito and Nakajima, Takashi and
                   Camenzind, Leon C and Kobayashi, Takashi and Scappucci,
                   Giordano and Tarucha, Seigo",
  month = aug,
  year = 2026,
  archivePrefix = "arXiv",
  primaryClass = "cond-mat.mes-hall",
  eprint = "2608.11072"
}

@article{muhonen_quantifying_2015,
  title = {Quantifying the quantum gate fidelity of single-atom spin qubits in silicon by randomized benchmarking},
  volume = {27},
  url = {http://dx.doi.org/10.1088/0953-8984/27/15/154205},
  DOI = {10.1088/0953-8984/27/15/154205},
  number = {15},
  journal = {Journal of Physics: Condensed Matter},
  publisher = {IOP Publishing},
  author = {Muhonen,  J T and Laucht,  A and Simmons,  S and Dehollain,  J P and Kalra,  R and Hudson,  F E and Freer,  S and Itoh,  K M and Jamieson,  D N and McCallum,  J C and Dzurak,  A S and Morello,  A},
  year = {2015},
  month = mar,
  pages = {154205}
}

@article{harper_statistical_2019,
  title = {Statistical analysis of randomized benchmarking},
  volume = {99},
  url = {http://dx.doi.org/10.1103/PhysRevA.99.052350},
  DOI = {10.1103/physreva.99.052350},
  number = {5},
  journal = {Physical Review A},
  publisher = {American Physical Society (APS)},
  author = {Harper,  Robin and Hincks,  Ian and Ferrie,  Chris and Flammia,  Steven T. and Wallman,  Joel J.},
  year = {2019},
  month = may
}

@ARTICLE{Rojas_Arias_2025_singleshotcorrelations,
  title = "Noise cross-correlations from single-shot measurements",
  author = "Rojas-Arias, J S and Stano, P and Wu, Y-H and Camenzind, L C and
               Tarucha, S and Loss, D",
  journal = {PRX Quantum},
  volume = {7},
  number = {2},
  pages = {020351},
  year = {2026},
  doi = {10.1103/1m1q-rvth}
}

@ARTICLE{Malinowski2017,
  title = "Spectrum of the nuclear environment for {GaAs} spin qubits",
  author = "Malinowski, Filip K and Martins, Frederico and Cywi{\'n}ski,
               {\L}ukasz and Rudner, Mark S and Nissen, Peter D and Fallahi,
               Saeed and Gardner, Geoffrey C and Manfra, Michael J and Marcus,
               Charles M and Kuemmeth, Ferdinand",
  journal = "Phys. Rev. Lett.",
  publisher = "American Physical Society (APS)",
  volume = 118,
  number = 17,
  pages = "177702",
  month = apr,
  year = 2017,
  doi = {10.1103/PhysRevLett.118.177702}
}

@ARTICLE{hrl_2026,
  title = "A digitally controlled silicon quantum processing unit",
  author = "{Members of the HRL Quantum Team and Collaborators}",
  journal = "Nature",
  publisher = "Springer Science and Business Media LLC",
  volume = 655,
  number = 8125,
  pages = "1154--1159",
  month = jul,
  year = 2026,
  doi = {10.1038/s41586-026-10754-7}
}

@ARTICLE{Aldeghi_nanomagnets_2025,
  title = {Simulation and Measurement of Stray Fields for the Manipulation of Spin Qubits in One- and Two-Dimensional Arrays},
  volume = {25},
  DOI = {10.1021/acs.nanolett.4c05037},
  number = {5},
  journal = {Nano Letters},
  publisher = {American Chemical Society (ACS)},
  author = {Aldeghi,  Michele and Allenspach,  Rolf and Vervelaki,  Andriani and Jetter,  Daniel and Bagani,  Kousik and Braakman,  Floris and Poggio,  Martino and Salis,  Gian},
  year = {2025},
  month = Jan,
  pages = {1838–1844}
}

@article{Unseld_baseband_2025,
  title = {Baseband control of single-electron silicon spin qubits in two dimensions},
  volume = {16},
  url = {http://dx.doi.org/10.1038/s41467-025-60351-x},
  DOI = {10.1038/s41467-025-60351-x},
  number = {1},
  journal = {Nature Communications},
  publisher = {Springer Science and Business Media LLC},
  author = {Unseld,  Florian K. and Undseth,  Brennan and Raymenants,  Eline and Matsumoto,  Yuta and de Snoo,  Sander L. and Karwal,  Saurabh and Pietx-Casas,  Oriol and Ivlev,  Alexander S. and Meyer,  Marcel and Sammak,  Amir and Veldhorst,  Menno and Scappucci,  Giordano and Vandersypen,  Lieven M. K.},
  year = {2025},
  month = Jul
}

@ARTICLE{Tadokoro_designs_2021,
  title = {Designs for a two-dimensional Si quantum dot array with spin qubit addressability},
  volume = {11},
  url = {http://dx.doi.org/10.1038/s41598-021-98212-4},
  DOI = {10.1038/s41598-021-98212-4},
  number = {1},
  journal = {Scientific Reports},
  publisher = {Springer Science and Business Media LLC},
  author = {Tadokoro,  Masahiro and Nakajima,  Takashi and Kobayashi,  Takashi and Takeda,  Kenta and Noiri,  Akito and Tomari,  Kaito and Yoneda,  Jun and Tarucha,  Seigo and Kodera,  Tetsuo},
  year = {2021},
  month = Sep
}

@ARTICLE{Bersano_magnetintegration_2025,
  title = {Integration of Cobalt Ferromagnetic Control Gates for Electrical and Magnetic Manipulation of Semiconductor Quantum Dots},
  volume = {35},
  url = {http://dx.doi.org/10.1002/adfm.202419940},
  DOI = {10.1002/adfm.202419940},
  number = {14},
  journal = {Advanced Functional Materials},
  publisher = {Wiley},
  author = {Bersano,  Fabio and Aldeghi,  Michele and Martinolli,  Niccolò and Boureau,  Victor and Aboud,  Thibault and Ghini,  Michele and Scarlino,  Pasquale and Salis,  Gian and Ionescu,  Adrian M.},
  year = {2025},
  month = Jan
}

@article{Struck2020,
  title = {Low-frequency spin qubit energy splitting noise in highly purified 28Si/SiGe},
  volume = {6},
  url = {http://dx.doi.org/10.1038/s41534-020-0276-2},
  DOI = {10.1038/s41534-020-0276-2},
  number = {1},
  journal = {npj Quantum Information},
  publisher = {Springer Science and Business Media LLC},
  author = {Struck,  Tom and Hollmann,  Arne and Schauer,  Floyd and Fedorets,  Olexiy and Schmidbauer,  Andreas and Sawano,  Kentarou and Riemann,  Helge and Abrosimov,  Nikolay V. and Cywiński, Lukasz and Bougeard,  Dominique and Schreiber,  Lars R.},
  year = {2020},
  month = May
}

@article{PaqueletWuetz2022,
  title = {Atomic fluctuations lifting the energy degeneracy in Si/SiGe quantum dots},
  volume = {13},
  url = {http://dx.doi.org/10.1038/s41467-022-35458-0},
  DOI = {10.1038/s41467-022-35458-0},
  number = {1},
  journal = {Nature Communications},
  publisher = {Springer Science and Business Media LLC},
  author = {Paquelet Wuetz,  Brian and Losert,  Merritt P. and Koelling,  Sebastian and Stehouwer,  Lucas E. A. and Zwerver,  Anne-Marije J. and Philips,  Stephan G. J. and Mądzik,  Mateusz T. and Xue,  Xiao and Zheng,  Guoji and Lodari,  Mario and Amitonov,  Sergey V. and Samkharadze,  Nodar and Sammak,  Amir and Vandersypen,  Lieven M. K. and Rahman,  Rajib and Coppersmith,  Susan N. and Moutanabbir,  Oussama and Friesen,  Mark and Scappucci,  Giordano},
  year = {2022},
  month = Dec
}

@article{DegliEsposti2024,
  title = {Low disorder and high valley splitting in silicon},
  volume = {10},
  url = {http://dx.doi.org/10.1038/s41534-024-00826-9},
  DOI = {10.1038/s41534-024-00826-9},
  number = {1},
  journal = {npj Quantum Information},
  publisher = {Springer Science and Business Media LLC},
  author = {Degli Esposti,  Davide and Stehouwer,  Lucas E. A. and G\"{u}l,  \"{O}nder and Samkharadze,  Nodar and Déprez,  Corentin and Meyer,  Marcel and Meijer,  Ilja N. and Tryputen,  Larysa and Karwal,  Saurabh and Botifoll,  Marc and Arbiol,  Jordi and Amitonov,  Sergey V. and Vandersypen,  Lieven M. K. and Sammak,  Amir and Veldhorst,  Menno and Scappucci,  Giordano},
  year = {2024},
  month = Mar
}

@misc{camenzind_intel_zenodo,
  author    = {Camenzind, L. C. and Jin, I. K. and Wu, Y.-H. and
               Noiri, A. and Takeda, K. and Nakajima, T. and
               Kobayashi, T. and Scappucci, G. and Tarucha, S.},
  title     = {{Supporting Data for ``Control--Noise Trade-offs in
                Industrially Integrated ${}^{28}$Si/SiGe
                Spin-1/2 Qubits''}},
  publisher = {Zenodo},
  year      = {2026},
  month     = oct,
  version   = {v1.0},
  doi       = {10.5281/zenodo.23584540},
  url       = {https://doi.org/10.5281/zenodo.23584540}
}

\end{document}